\pdfoutput=1
\documentclass[nodraft,structabstract]{aa}  
\usepackage[T1]{fontenc}
\usepackage{aecompl}

\usepackage{natbib}
\usepackage{graphicx}
\usepackage{float}
\usepackage{url}
\usepackage{txfonts}
\newcommand{\degree}{$^\circ$}
\newcommand{\HA}{H$\alpha$}

\begin{document}
   \title{Study of FK Comae Berenices.}

   \titlerunning{Study of FK Comae Berenices. VII.}

   \subtitle{VII. Correlating photospheric and chromospheric activity\thanks{Based on the observations obtained at the Nordic Optical Telescope, Observatorio Roque de los Muchachos, La Palma, Canary Islands, Spain; Kitt Peak National Observatory, USA}}

   \author{K. Vida\inst{1}
          \and
          H. Korhonen\inst{2, 3}
          \and
          I.V. Ilyin\inst{4}
          \and
          K. Ol\'ah\inst{1}
          \and
          M.I. Andersen\inst{3, 5}
          \and
          T. Hackman\inst{6}
          }
   \institute{
	   Konkoly Observatory, MTA CSFK, H-1121 Budapest, Konkoly Thege M. út 15-17, Hungary\\ 
     \email{vidakris@konkoly.hu}
     \and
     Finnish Centre for Astronomy with ESO (FINCA), University of Turku, V\"ais\"al\"antie 20, FI-21500 Piikki\"o, Finland \\\email{heidi.h.korhonen@utu.fi}
     \and Centre for Star and Planet Formation, Niels Bohr Institute, University of Copenhagen, {\O}ster Voldgade 5-7, DK-1350, Copenhagen K, Denmark
     \and Leibniz-Institut f{\"u}r Astrophysik Potsdam (AIP), An der Sternwarte 16, D-14482 Potsdam, Germany 
     \and Dark Cosmology Centre, Niels Bohr Institute, University of Copenhagen, Juliane Maries Vej 30, DK-2100 Copenhagen {\O}, Denmark
     \and Department of Physics, PO Box 64, FI-00014 University of Helsinki, Finland}

   \date{Received September 15, 1996; accepted March 16, 1997}

% \abstract{}{}{}{}{} 
% 5 {} token are mandatory
 
  \abstract
   {}
   {We study the connection between the chromospheric and photospheric behaviour of the active late-type star FK\,Comae.
	   }
   {We use spot temperature modelling, light curve inversion based on narrow- and wide-band photometric measurements, \HA{} observations from 1997--2010, and Doppler maps from 2004--2010 to compare the behaviour of chromospheric and photospheric features. }
   { Investigating low-resolution \HA{} spectra we find that the changes in the chromosphere seem to happen mainly on a time scale longer than a few hours, but shorter variations were also observed. According to the \HA{} measurements prominences are often found in the chromosphere that reach to more than a stellar radius and are stable for weeks, and which seem to be often, but not every time connected with dark photospheric spots. The rotational modulation of the \HA{} emission seems to  typically be anticorrelated with the light curve, but we did not find convincing evidence of a clear connection in the long-term trends of the \HA{} emission and the  brightness of the star. In addition, FK\,Com seems to be in an unusually quiet state in 2009--2010 with very little chromospheric activity and low spot contrast, that might indicate the long-term decrease of activity.}
   {}

   \keywords{stars: activity --  stars: magnetic field --  stars: atmospheres --  stars: chromospheres --  stars: late-type --   stars:individual:FK\,Com}

   \maketitle
%
%________________________________________________________________

\section{Introduction}

FK Comae is the eponymous member of a group of active stars \citep{1981IAUS...93..177B}, where the members are fast-rotating G--K giants showing chromospheric and coronal activity signs similar to the RS~CVn-type binaries. Their typical $v\sin i$ is of order of 100km\,s$^{-1}$, but no periodic radial velocity variations have been detected, indicating that these stars are single, or have a companion of very low mass. 
Studying the Na D line, \cite{1984ApJ...283..200M} gave 5\,km\,s$^{-1}$ as an upper limit of the radial velocity variations caused by a possible companion, corresponding to a mass of 0.054$M_{Sun}$. \cite{Huenemoerder} gave an upper limit of the semi-amplitude of 3\,km\,s$^{-1}$, which makes the presence of the companion even more unlikely.
Photometric observations showed periodic, quasi-sinusoidal changes of $\Delta V=0\fm1-0\fm3$ interpreted as starspots by \cite{1981IAUS...93..177B}. \cite{1981IAUS...93..177B} also suggested,  based on the rapid rotation and the lack of radial velocity variations, that these stars are coalesced W~UMa-type systems following the scenario proposed by \cite{Webbink:1976db}.

FK Com is the most active, and most thoroughly studied star of this group. The star shows signs of activity from photosphere to corona \citep{1948PASP...60..382M,1966IBVS..172....1C,1981ApJ...251L.101R,1984PASP...96..250D,1999A&A...343..213O, ayres, drake}. FK Com was also the first object where the so-called flip-flop phenomenon was observed \citep{1993A&A...278..449J,1994A&A...282L...9J}: the star shows active longitudes separated by $\sim$180\degree, and the dominance between these regions is exchanged  every few years. \cite{fkcom6} found that both flip-flops and phase jumps can be found on FK Com. They concluded, that these are two different, although possibly connected phenomena. \cite{2013A&A...553A..40H} also showed that flip-flops are not a single phenomenon: they can occur gradually (through differential rotation) or suddenly, the phase shift varies and also as does the stability of the new primary spot region.

\cite{1981ApJ...251L.101R} described the broad, double-peaked \HA{} profiles of FK Com with an ``excretion'' disk around the star, which itself has a small contribution to the \HA{} emission. \cite{Huenemoerder} presented photometric and spectroscopic data and concluded that FK Com is a single star with extended matter around it. 
From long-term \HA{} observations \cite{1993PASP..105.1427W} found evidence of co-rotating emitting material around FK Com within two stellar radii with lifetimes longer than a few weeks.
\cite{1999A&A...343..213O} modelled highly variable Balmer line profiles and confirmed the complex circumstellar structures. \cite{2005A&A...434..221K} presented high-resolution \HA{} observations spanning over two years. The authors suggested a model where FK Com is a binary system with a low-mass secondary star illuminating a circumstellar accretion disk. The model contains additional sources of absorption and emission at  phases 0.0 and 0.5, respectively. 

In this paper, the seventh in a series on FK Comae, we analyse \HA{} data  spanning over a decade from 1997 to 2007. 
The chromospheric activity patterns from the H$\alpha$ data are compared, in most epochs, to the contemporaneous photospheric spot maps obtained using the Doppler imaging technique.

\section{Observations}
\begin{table}
\caption{List of \HA{} observations}
\centering
  \begin{tabular}{cccc}
\hline\hline
\multicolumn{4}{c}{High-resolution spectra}\\
Date& Telescope &\# of spectra\\
\hline
1997/04/07--04/16& KPNO& 10&\\ 
1998/03/10--03/17& NOT &  7&\\
1998/04/01--04/22& KPNO& 19&\\ 
1998/07/02--07/15& NOT & 38\\
1999/03/03--03/05& NOT &  6&\\ 
1999/05/24--06/04& NOT & 19&\\ 
1999/07/23--08/03& NOT & 12&\\ 
2000/03/29--04/08& KPNO& 10&\\
2000/04/26--05/05& KPNO& 10&\\
2000/08/07--08/18& NOT & 12&\\ 
2001/05/06--05/09& NOT &  6&\\ 
2001/06/03--06/12& NOT & 17&\\ 
2002/08/20--08/29& NOT &  7&\\ 
2003/06/03--06/22& NOT & 13&\\ 
2004/01/30--02/11& NOT & 10&\\ 
2004/07/25--08/08& NOT & 10&\\ 
2005/07/15--07/23& NOT &  7&\\ 
2007/07/18--07/29& NOT & 12&\\ 
2009/08/26--09/06& NOT & 11&\\
2009/12/27--01/04& NOT & 18&\\
\hline
\smallskip\\
\multicolumn{4}{c}{Low-resolution spectra}\\
Date& Telescope &\# of spectra&Phase\\
\hline
1999/02/06 &NOT &11&0.35--0.40\\
1999/02/07 &NOT &43&0.75--0.78\\
\hline
  \end{tabular} 
\label{tab:obslog}
\end{table}
%----
Most of the spectroscopic observations presented here were obtained at the 2.56\,m Nordic Optical Telescope (NOT) during 13 observing runs using the SOFIN high resolution spectrograph and one run of low resolution spectroscopy using ALFOSC. In addition, three datasets of high resolution spectra were obtained at the Kitt Peak National Observatory (KPNO) in Arizona. All the runs from KPNO and 10 runs with the SOFIN at NOT (1998--2003 and 2004 February) were already used in \cite{fkcom6}, \cite{ayres} and \cite{fkcom5} for investigating the photospheric activity. Thus, here only the unpublished observations from SOFIN and ALFOSC are described in detail. In addition to these spectroscopic observations, also previously published photometric observations from \cite{fkcom3} and \cite{2013A&A...553A..40H} are used.

The phases for all the observations presented here were calculated using the ephemeris obtained from 25 years of photometric observations, $\mathrm{HJD}= 2,439,252.895 + (2\fd4002466\pm0\fd0000056)E$, referring to a photometric minimum calculated by \cite{1993A&A...278..449J,1994A&A...282L...9J}.

\subsection{High resolution spectroscopy}
\begin{figure*}[t]
 \centering
 \includegraphics[angle=-90,width=0.9\textwidth]{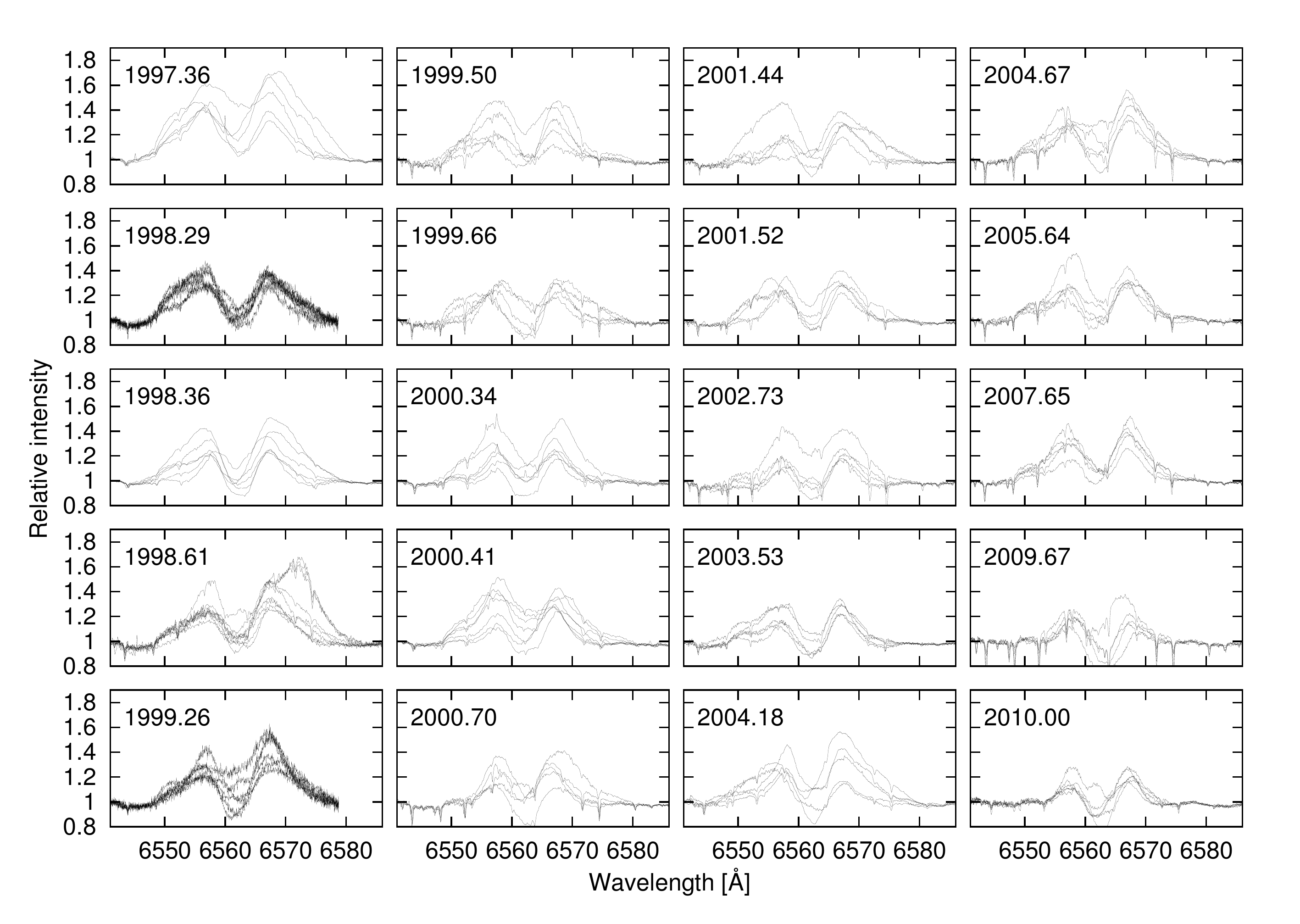}
 \caption{Samples of high-resolution \HA{} profiles from different observing runs.}
 \label{fig:spec_sample}
\end{figure*}
%------
High resolution NOT observations for  five epochs (2004 July, 2005 July, 2007 July, 2009 August and 2010 January) were carried out using SOFIN \'echelle spectrograph with the low resolution camera, giving resolving power ($\lambda/\Delta\lambda$) of 27,000. The \'echelle spectra consisted of typically 45 orders and were centred at 5200~{\AA}. The signal-to-noise ratio in the H$\alpha$ order varied between  90 and 310, typical value being around 200. The data were reduced using the 4A reduction package \citep{ilyin}. Table~\ref{tab:obslog} gives the summary of the high resolution spectra used in this paper, including the previously published datasets.

\subsection{Low resolution spectroscopy}

For investigating the possible rapid variations in the H$\alpha$ profiles, low resolution spectra with high cadence were obtained using ALFOSC at the NOT during 1999 February 26--27. We used the 0.4 arcsec slit and the \'echelle grism \#9 together with the H$\alpha$ filter \#22. With this configuration we selected only the order containing H$\alpha$, without using a cross disperser. The resolving power obtained with this configurations was approximately 4,000. 

The observations were done in sets of two observations with 60 sec exposure times, and taking Halogen flat and Helium--Neon arc observations before and after each pair. Due to selecting only a narrow spectral region only a couple of arc lines were seen. Thus, no accurate wavelength calibration could be carried out. An approximate wavelength solution was obtained from the theoretical calculations for the grism \#9, and the solution was fine-tuned for each observation by cross-correlating the arc spectra and introducing the shifts obtained to corresponding observation. As we were only interested in the rapid variability of the profiles this procedure gave a wavelength solution that was sufficient for our purpose. All the observations were reduced using the Image Reduction and Analysis Facility (IRAF) which is distributed by KPNO/NOAO.

\subsection{Photometry}

In this paper beside the $\Delta V$ observation presented in \cite{cycles} and \cite{2013A&A...553A..40H}, we use the all the data taken in different other colours ($U, B, I_C, b, y$) parallel with the $V$ measurements of \cite{fkcom3, fkcom5}, to estimate the temperatures of active regions in FK~Com.

\section{Results}

\subsection{Long term chromospheric activity}
\begin{figure*}
 \centering
 \includegraphics[width=0.85\textwidth]{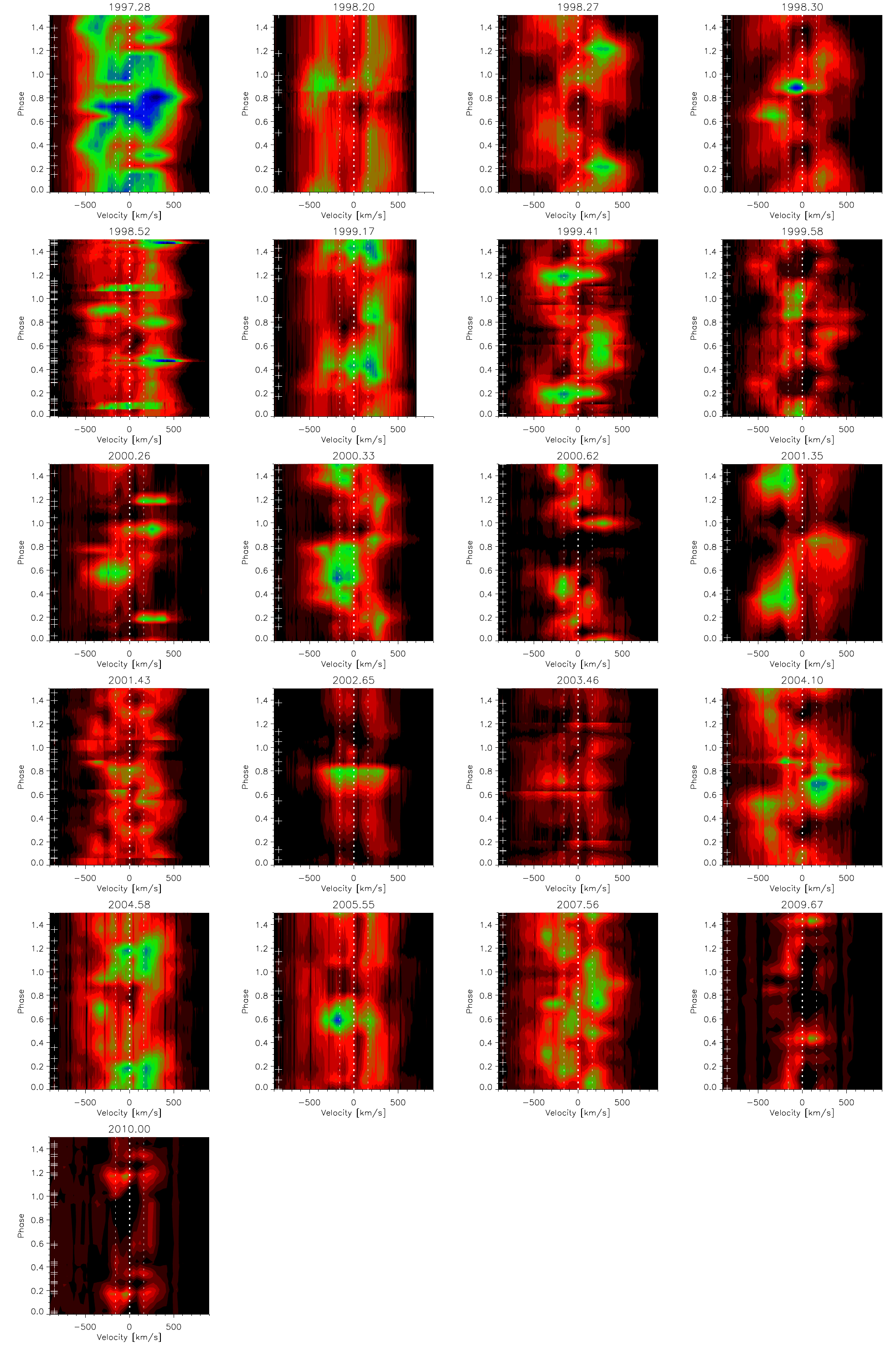}
 \caption{Dynamic H$\alpha$ residual spectra of FK\,Comae. Crosses on the left show the phases of individual spectra. Thick dashed lines mark the 0 \,km\,s$^{-1}$ velocity shift (i.e., the centre of the stellar disk), the thin lines correspond to the edges of the disk. }
 \label{fig:dynamic-spectra}
\end{figure*}
\begin{figure}
 \centering
 \includegraphics[width=0.44\textwidth]{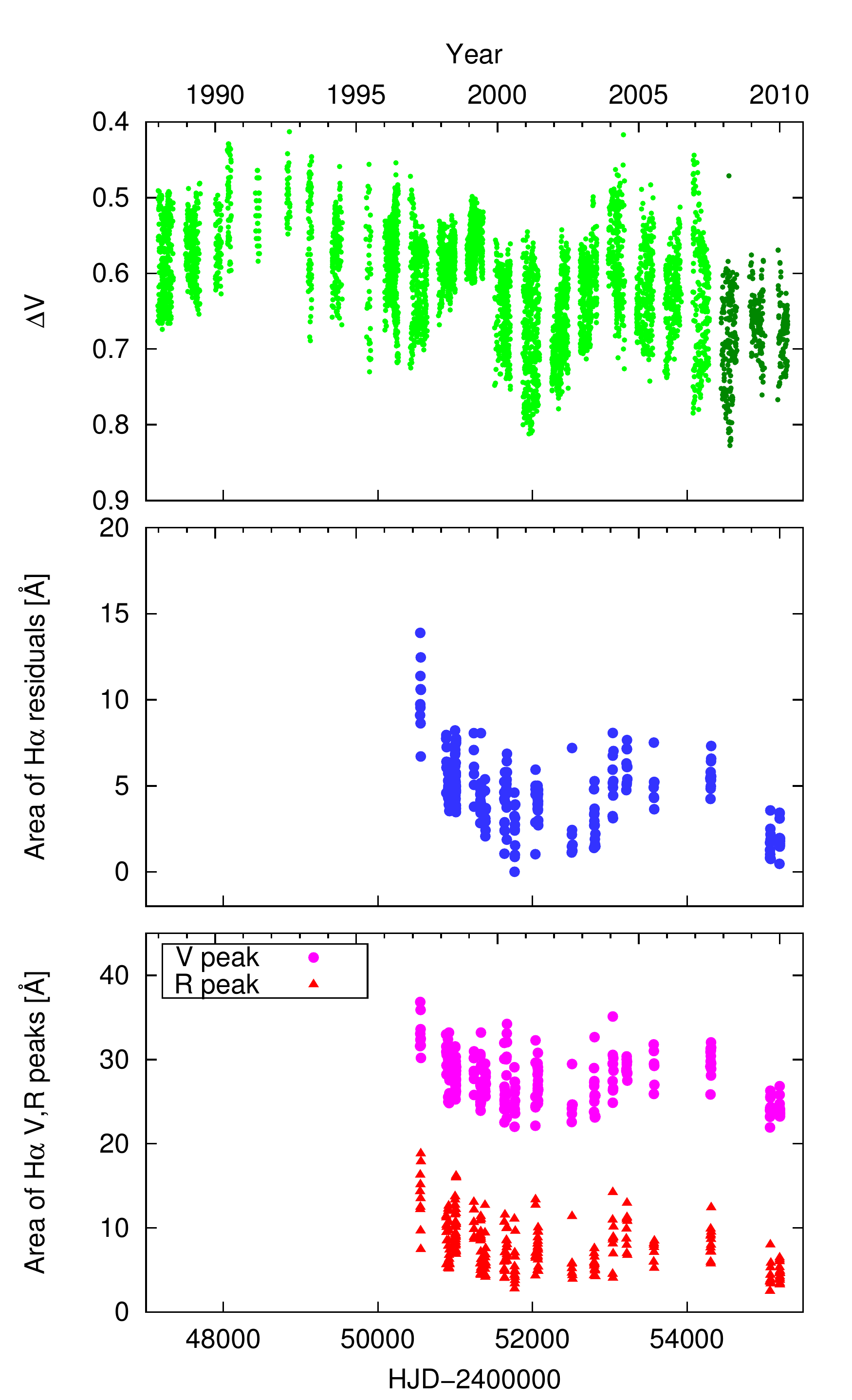}
 \caption{From top to bottom: $\Delta V$ photometry of FK\,Comae from \cite{cycles} and \cite{2013A&A...553A..40H} (the latter after 2008, with darker colour), \HA{} line areas from Gaussian fits of
  the residual spectra, and the area change of the \HA{} violet (V) and red (R) peaks (V peak shifted arbitrarily by 20\,\AA{}  for clarity).  
  }
 \label{fig:ha-strength}
\end{figure}
A sample of the \HA{} spectra is plotted on Fig. \ref{fig:spec_sample}. For each observing run 5--10 spectra that show well the behaviour of the region were chosen and plotted. As can be seen, the profiles are highly variable, but at each epoch 
 display a broad double-peaked emission which is indicative of a circumstellar disc. 

The complex \HA{} profile shape of FK\,Com indicates many different components: circumstellar disc, plages and prominences. To study just the contribution from the stellar component (plages and prominences) the disc contribution is removed. To facilitate this, we  subtracted the spectrum that showed the least amount of stellar \HA{} emission and seemed to only show the  circumstellar disc component (a NOT spectrum from 2000/08/10) having a symmetric \HA{} profile with no other emission feature than the two circumstellar disc-peaks. However, it is possible that this lowest intensity spectrum is just a local activity minimum in the decade-long observations and even this one contains some low-level plage- or prominence-like chromospheric activity. Using these residual spectra, we have plotted the dynamic spectra on Fig. \ref{fig:dynamic-spectra} for each observing run.  The plots show the excess emission at different velocities (x-axis) and rotational phased (y-axis, times of observations marked with plus-signs). The dashed white line denotes the location of the stellar disc centre and the dotted lines the edges of the stellar disc. Colour coding (black--red--green--blue) shows the intensity of the \HA{} region. Crosses mark the phases of the observations, for phases where there is no data the plot is interpolated to the closest measurements. More information on how these  dynamical spectra have been calculated can be found from \cite{2009MNRAS.395..282K}. On the whole, most of the excess emission is seen outside the stellar disc, implying material located higher in the stellar atmosphere. Sometimes, excess emission is also seen on the stellar disc, implying plage-like region in the chromosphere of FK\,Com. The wave-like features seen on these plots  are probably the result of prominences rotating well outside the stellar radius. This feature can be best seen at 1998.27, 1998.30, 1999.41, 2000.33, 2001.35, 2004.10, 2004.58, and 2007.56. The shape of the waves is continuously changing, indicating that these prominences are constantly evolving. 
It is possible, that some of the prominence-like features is related to the circumstellar disc: the disc could get heated unevenly, for example by stellar flares. This could cause some transient features in the disc emission. We cannot completely rule out that some of the emission seen would come from the disc, but the activity of the star itself is more likely the origin (the star is very active, as shown by all indicators).

To check the long-term behaviour of the \HA{} line, and give a more quantitative description, we fitted Gaussians to the residual spectra for measuring the changes of the whole \HA{} region, and a simple model consisting of two Gaussians to the original spectra  to describe the violet and red peaks of the \HA{} region individually. In Fig. \ref{fig:ha-strength} we plotted the $\Delta V$ light curve from \cite{cycles} and \cite{2013A&A...553A..40H} together with the strength of the \HA{} emission measured as the area of the fitted Gaussians (both from the residuals and the individually fitted violet and red peaks). 
The strength of the line decreases until $\sim$2002--2003, then increases in all cases, indicating a change of chromospheric activity on a time scale of a few years. 

\subsection{Fast variability in chromospheric activity}
\begin{figure}
 \centering
 \includegraphics[angle=-90,width=0.33\textwidth, angle=90]{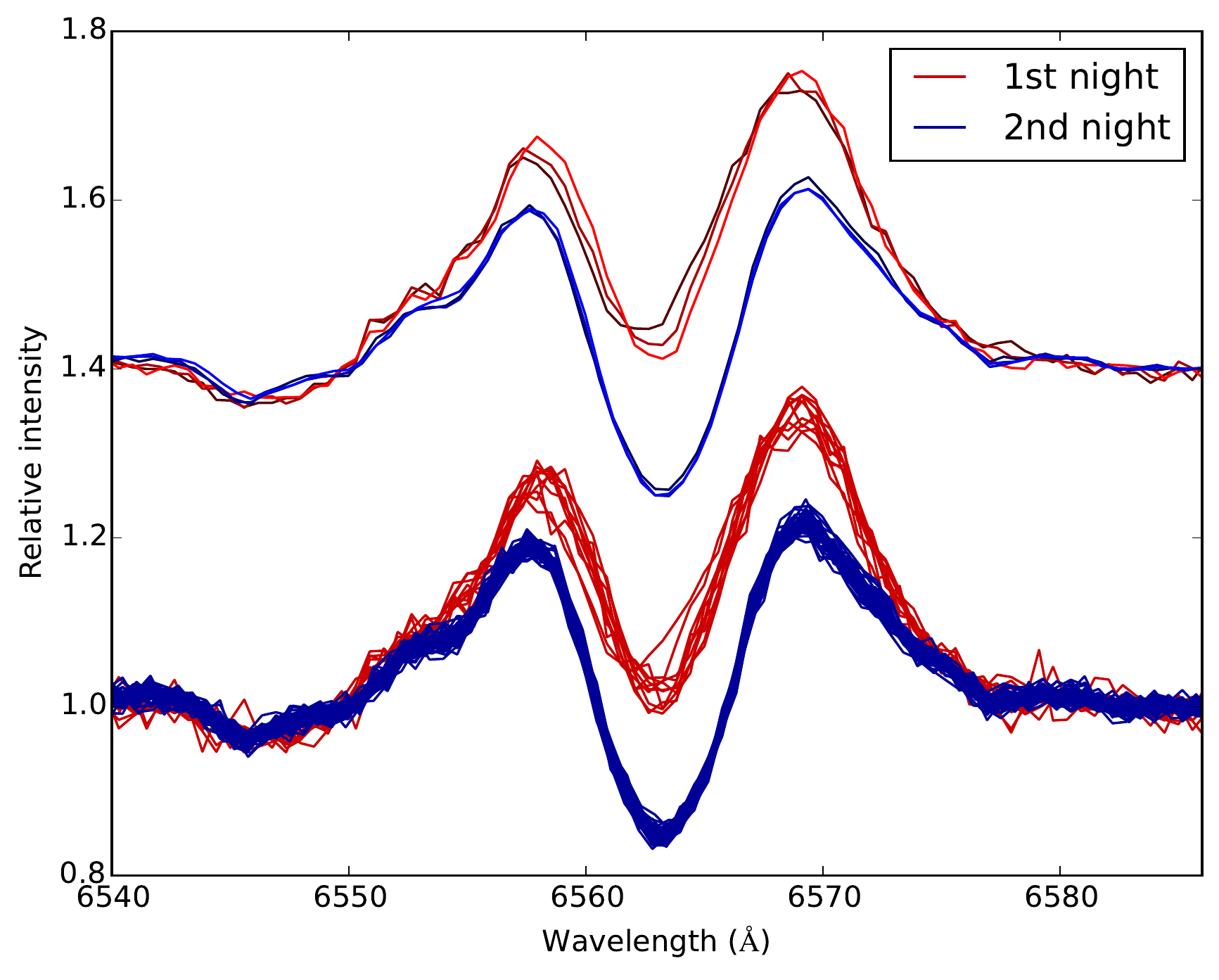}
 \caption{Low-resolution \HA{} from 1999 February. Red continuous line shows the first, blue dashed line shows the second night. The first night covers $\sim3$, the second night $\sim2$ hours.
 The same spectra, averaged in 30 min windows, are plotted above the original, with colours brightening with time. 
 }
 \label{fig:lowres}
\end{figure}

Along with the long-term variations, we have also searched for minute scale variations using lower resolution spectra obtained in 1999 February (see Table \ref{tab:obslog}). The spectra from the two nights are plotted in Fig. \ref{fig:lowres}. 
The \HA{} profile of FK\,Com  does not show relevant changes during the time span of about 2--3 hours of each night.
The observations from the first night show more small scale variations.  To check if these changes are real or caused by the lower signal-to-noise ratio of these observations (average S/N was 88 on the first night and 118 on the second), we averaged the spectra in 30-minute windows, and plotted the result in Fig. \ref{fig:lowres}. These averages indicate small variations in the \HA{} region on the time scale of hours, especially during the first night which is exhibiting higher activity levels. 
Anyhow, there is more prominent difference between the two nights. Thus we can conclude that, at least during the two nights  of our  observing run, the chromospheric structure of the star shows significant night-to-night variability, and much weaker variations on smaller time scales.
Note also that the low resolution of the observations can mask some  smaller scale variability.

\subsection{Photospheric spots from Doppler imaging}

\begin{figure*}
 \centering
 \includegraphics{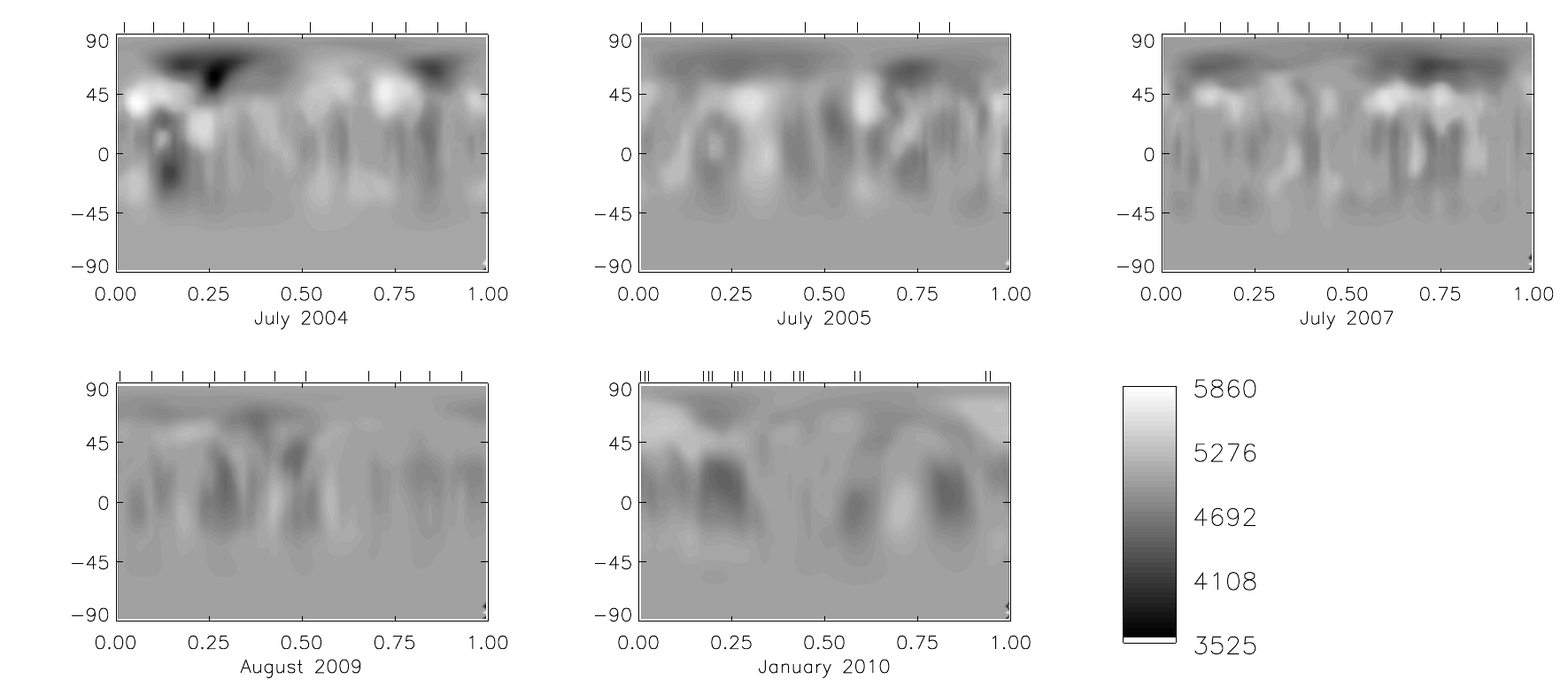}
 \caption{The surface temperature maps for 2004 July, 2005 July, 2007 July, 2009 August and 2010 January datasets. The lines above the maps mark the phases of the observations. }
 \label{fig:maps}
\end{figure*}

For the five previously unpublished high resolution spectral datasets we also obtain surface temperature maps using Doppler imaging techniques. For this the Tikhonov Regularisation inversion code INVERS7PD, written by N.\ Piskunov \citep{piskunov90} and modified by T.\ Hackman \citep{hack} was used. The adopted stellar parameters and additional details on the inversion technique and line-profile calculations can be found,  e.g.,  in \cite{fkcom2, fkcom5}. The obtained temperature maps are shown in Fig.~\ref{fig:maps} and discussed in the following.

In 2004 July observations from 10 rotational phases of FK Com were obtained. The phase coverage of these observations is good, and the largest phase gap is 0.17, spanning the phases 0.36--0.52. The map obtained from the data shows two high latitude,  47--79$^{\circ}$, spot concentrations around the phases 0.25 and 0.85. A lower latitude active region is seen around the phase 0.15 spanning the latitudes 20--29$^{\circ}$. The average spot latitude obtained from this map is $54.9\pm 3.4^{\circ}$ according to the output of the inversion code. The average spot latitudes were determined from the Doppler images using areas that were cooler than the limiting temperature of $T_\mathrm{lim}=4550$\,K ($T_\mathrm{eff}$ = 5000\,K unspotted surface temperature assumed), its error was estimated by the standard error of the weighted mean. The area at individual latitude strips were measured, and in the averaging the spot areas at each latitude were used as weights. The two spots located at the phases 0.1--0.3 are approximately 1200~K cooler than the unspotted surface, and the one at the phase 0.8 is 800~K cooler. The coolest temperatures in the spot group around the phases 0.1--0.3 are up to 1500~K cooler than that of the unspotted surface.

The SOFIN observations obtained in 2005 July consist of eight different phases, with the largest gap spanning the phases 0.17--0.45. Not many cool regions are seen in this epoch. Only one spot, which is approximately 500~K cooler than the unspotted surface, is present at phases 0.7--0.8 spanning the latitude range 56--70$^{\circ}$. The average latitude of this region is $61.1\pm 0.7^{\circ}$. The coolest temperature in this spot region is about 600~K cooler than that of the unspotted surface.

The phase coverage of the 12 spectra obtained in 2007 July is very good and there are no phase gaps larger than 0.1. Also this map shows, similarly to the 2004 July one, two spot concentrations on the surface. The main spot group is approximately 600~K cooler than the unspotted surface, with the coolest temperatures going down to 800~K cooler than the unspotted surface. This spot is located at the phases 0.7--0.8 and spans the latitudes 56--74$^{\circ}$. The second spot group has similar temperatures and it is located at phases 0.1--0.2 and spans latitudes 61--70$^{\circ}$. The average spot latitude is $65.3\pm 0.8^{\circ}$.

The observations carried out in 2009 August consist of 11 spectra, which are relatively evenly distributed over the stellar rotation. The largest phase gap of 0.17 in phase is located at the phases 0.51--0.68. Similarly to the July 2005 map there are no prominent cool spots at this epoch. The coolest regions, which occur around phases 0.25--0.5, are only approximately 500~K less than the temperature of the unspotted surface. The latitude of the spots is also relatively low compared to what is typically seen in FK~Com, 7--34$^{\circ}$ with and average latitude of $18.5\pm 1.5^{\circ}$.

In 2010 January 18 spectra were obtained on three nights during a four night time period. Therefore, even though there are many phases, they are grouped together and there are large phase gaps in the dataset. The largest gap occurs between the phases 0.59 and 0.93. As in the August 2009 and July 2005 map the cool spots are of very low contrast and the difference between them and the unspotted surface is only approximately 550~K. Two main spot groups can be seen on the surface, one spanning phases 0.15--0.25 and another one around phase 0.8. As is the case for August 2009, also in January 2010 the spots are located at low latitudes, 2--25$^{\circ}$ with and  average latitude of  $12.4\pm 1.2^{\circ}$.

\section{Discussion}
\label{sect:discussion}

\subsection{Photospheric activity}
\begin{figure*}
 \centering
 \includegraphics{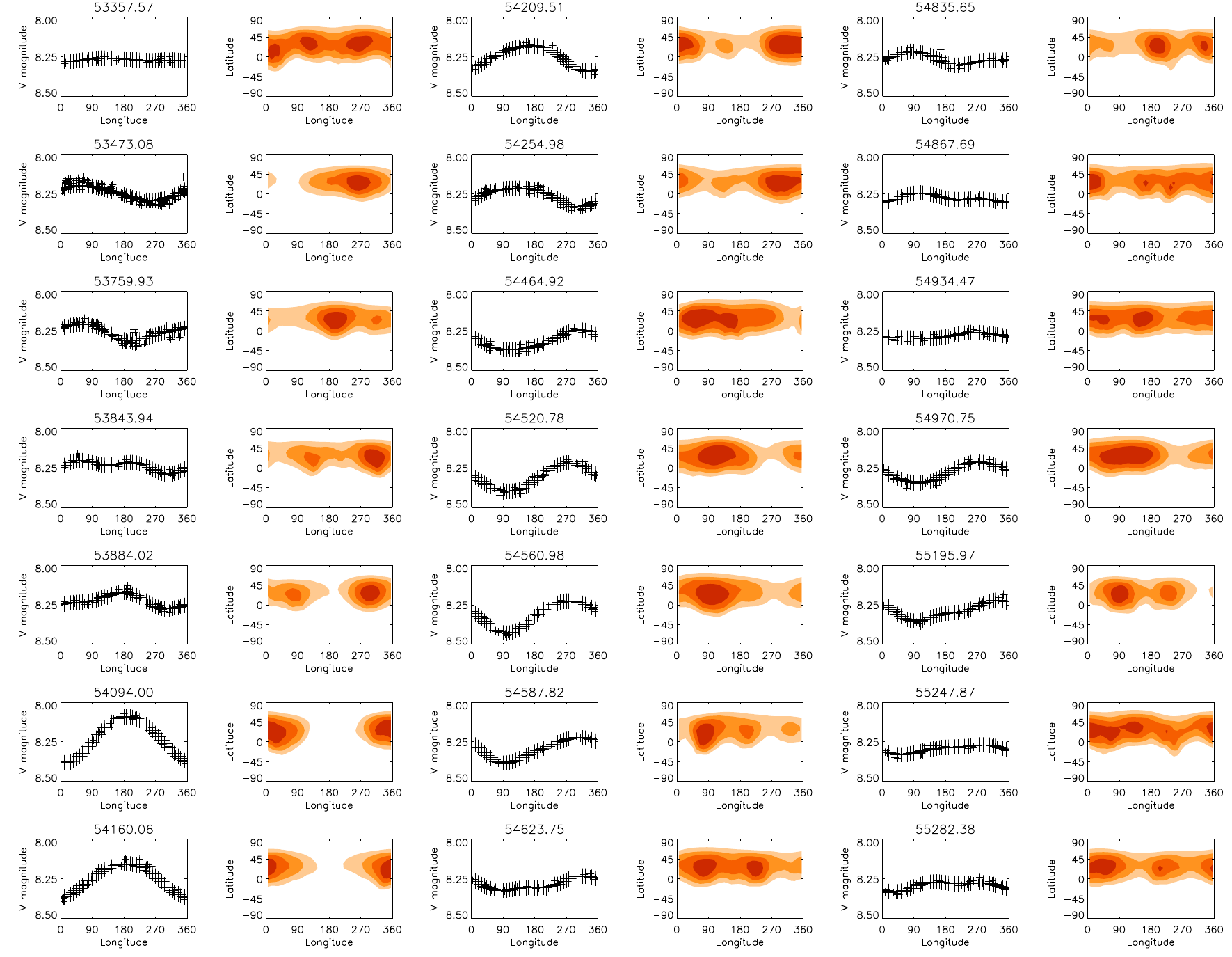}
 \caption{The spot filling factor maps for 2005 January -- 2010 April. In light curve plots the crosses denote the observations and the solid line the inversion result. In the spot filling factor maps the darker colour means larger spot filling-factor. In both plots the x-axis is longitude in degrees and the y-axis in the light curve plots is V magnitude and in the spot filling factor maps latitude in degrees. The mean Julian Date (HJD$-2400000$) of the observations for each set is given on top of the light curve plots. }
 \label{fig:LI}
\end{figure*}

The evolution of the photospheric activity of FK Com in 2004--2010 can be studied based on the five new Doppler images presented in this paper.

The first map, July 2004, has the highest spot contrast of all the new maps presented here. Second highest spot contrast is seen in the July 2007 map, where a spot contrast typical of FK~Com of approximately 800~K is seen. July 2005, August 2009, and January 2010 maps, on the other hand, have similarly low spot contrasts with the coolest spots being only 500--600~K cooler than the unspotted surface. Often this kind of low contrast maps are produced if the data quality is bad, but here especially the July 2007 and August 2009 maps are of good quality, and particularly have very good phase coverages. Therefore, it is clear that the temperature of the spots has drastically changed between July 2004 and July 2005, and that also in the latter years FK~Com was still in the low activity state. Note, that hot features in the maps at the same longitude as the prominent cool spots are most likely artefacts caused by changing continuum levels with changing temperature. If the hot features are well separate from the cool spots they could be real (see Section \ref{sect:general}).

All the maps show the spots at similar phases, around 0.2 and 0.8. All the spots are of very low contrast in the July 2005, August 2009 and January 2010 maps and it is debatable if they can even be considered active regions. The spot group around phase 0.8--0.9 is seen in all the maps, except August 2009, but in the July 2004 map it is the secondary spot group and the main group in the July 2005 and the July 2007 maps. This could indicate a possible flip-flop event between July 2004 and July 2005, an event which is supported by the July 2007 map. On the other hand, the two later maps, August 2009 and January 2010, again show 0.2 as the more active phase. This would imply yet another flip-flop, which would have taken place between 2007 and 2009. \cite{fkcom4} reported that flip-flop events occur on average every 3.2 years, which would mean that two flip-flops could in principle have happened in the 5.5 years the observations cover (although the flip-flops are not regular phenomena, see e.g. \citealt{fkcom6, 2013A&A...553A..40H}). Still, as is also concluded by \cite{fkcom5}, flip-flops are difficult to detect from sparse Doppler imaging alone, and are better seen based on photometric observations with denser coverage. For this reason we also carried out light curve inversion, where inversion techniques are applied to broad-band photometry and maps of the fraction of each pixel on stellar surface covered by spots, so-called spot filling factor, are calculated. More on the technique can be found, e.g., from \cite{lanza}.

The light curve inversion results are shown in Fig.~\ref{fig:LI}. The data were divided into 21 subsets based on the stability of the light curve. Continuous phase shifts of the spotted longitudes can be followed in 6 subsequent maps between the JD 2453843.94--2454254.98 maps with one seasonal gap of 150 days in the data. Another continuous shifting of the spots is observed on 4 consecutive maps between  2454520.78 and 2454623.75, corresponding to data without long interruptions. Drastic difference is seen between the maps of 2454254.98 and 2454464.92, and 2454623.75 and 2454835.65, with  seasonal gaps in the data of about 150 days. These cases, in lack of more continuous sequences, the changes can also be explained by continuous phase shifts, but could not be followed. However, a substantial change of the activity pattern, although shown by low amplitude light curves, happened between 2454867.69 and 2454934.47 maps, with a small gap of 28 days in the data, i.e., sometime in 2009 March, which can be considered as a flip-flop-like event (see the Discussion later).

The spot filling factor maps can be compared with the Doppler images. The spot filling factor map closest to the July 2005 is the one from Julian Date 2453473.08 (second map in the plot). The location of the main spot measured from the spot filling factor map is 0.73 in phase, similarly to what is seen in the Doppler image. The July 2007 Doppler image is close to the spot filling factor map from Julian Date 2454254.98. Again the location of the main spot in both maps is very similar. For August 2009 there is no contemporaneous spot filling factor map, but the map of 2454970.75 is obtained few months before it. The Doppler image of August 2009 shows very little spots, but they are located at same phases as the main activity seen in the spot filling factor map. Naturally, the activity could have gotten weaker within the approximately two months separating the spot filling factor map and the Doppler image. The spot filling factor map of Julian Date 2455247.87 is contemporaneous to the January 2010 Doppler image. The spot filling factor map, like the Doppler image, shows main activity around the phase 0.25, and the surface  could be covered by relatively low contrast spots.

On the whole the spot filling factor maps obtained from the photometry correspond very well to the Doppler images that are based on high resolution spectra. The possible flip-flop events seen in the Doppler images are also seen in the spot filling factor maps, but suspiciously most of them occur during the long gap when the star is not observable.

\subsection{Correlating chromospheric and photospheric activity regions}
\begin{figure}
 \centering
 \includegraphics[width=0.16\textwidth]{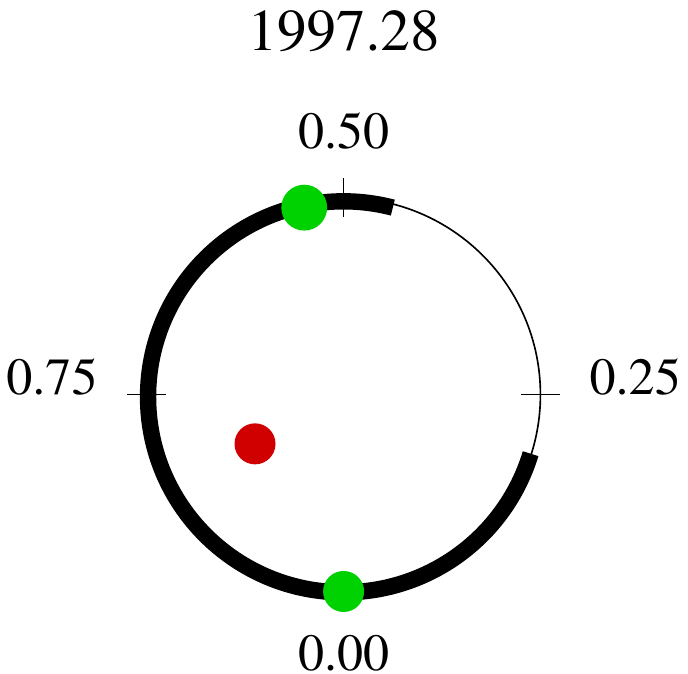}
 \includegraphics[width=0.16\textwidth]{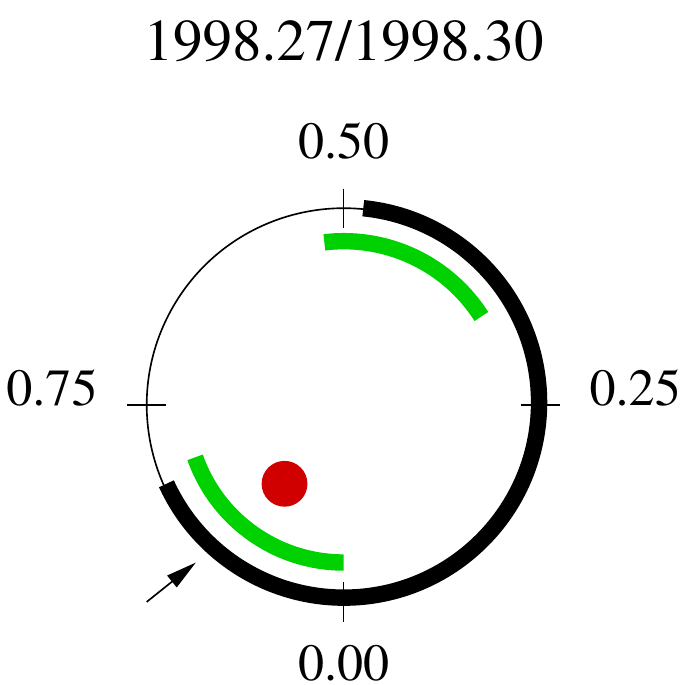}
 \includegraphics[width=0.16\textwidth]{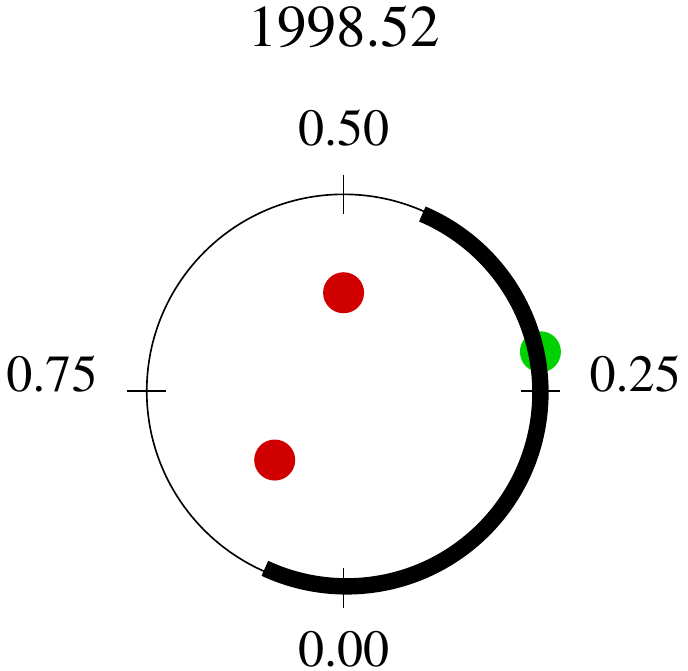}\\
\vspace{6mm}
 \includegraphics[width=0.16\textwidth]{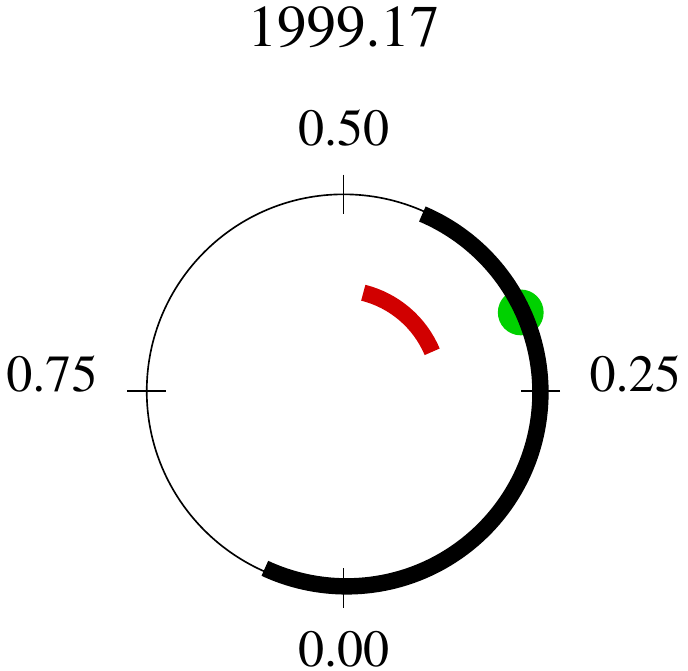}
 \includegraphics[width=0.16\textwidth]{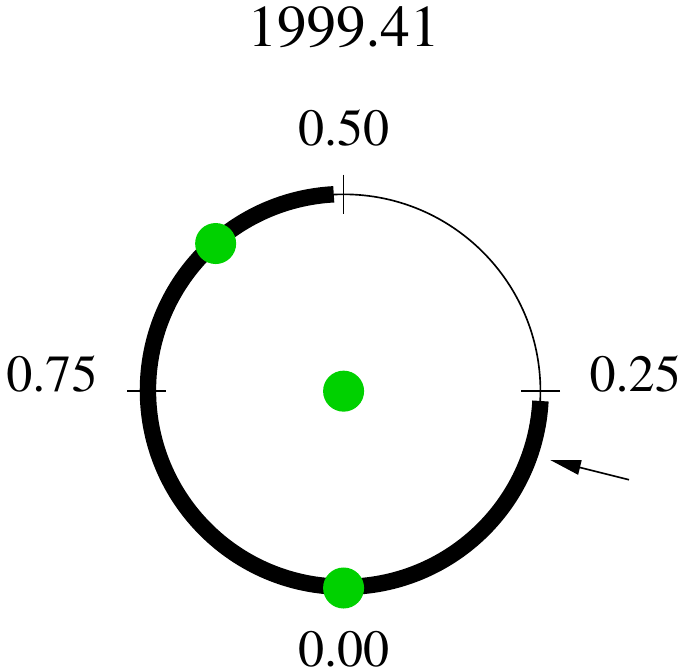}
 \includegraphics[width=0.16\textwidth]{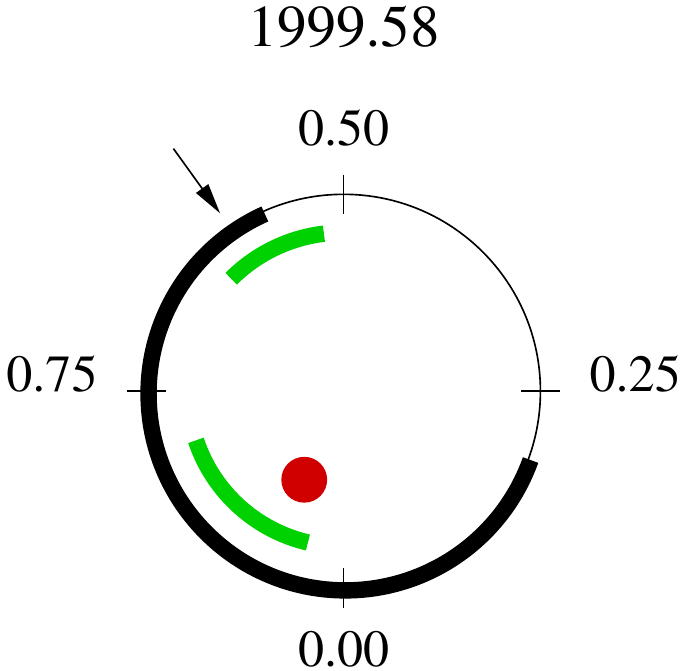}\\
\vspace{6mm}
 \includegraphics[width=0.16\textwidth]{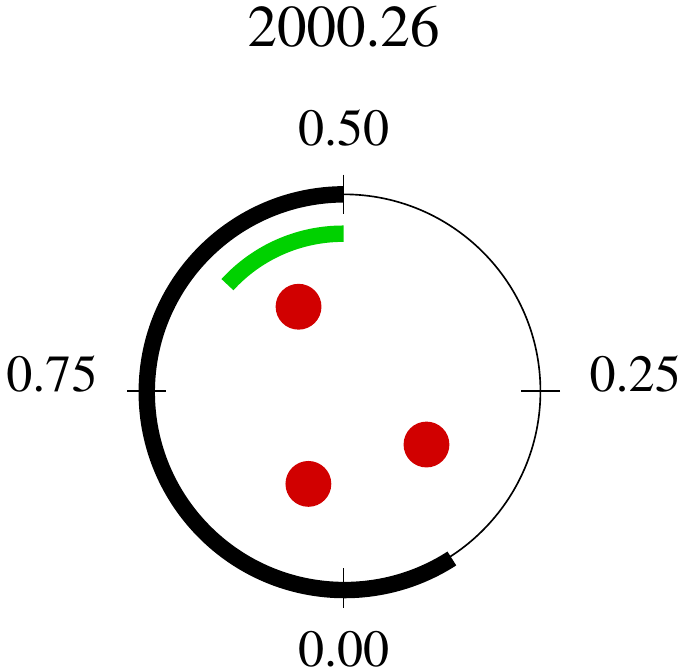}
 \includegraphics[width=0.16\textwidth]{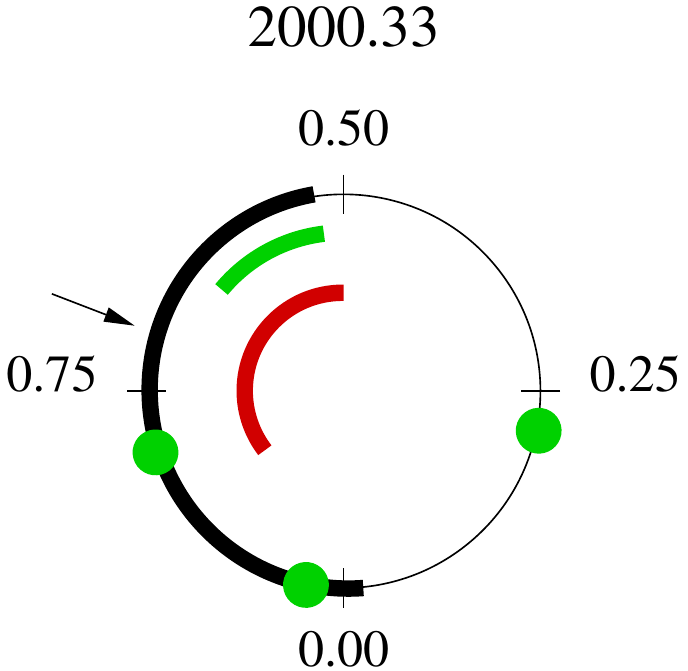}
 \includegraphics[width=0.16\textwidth]{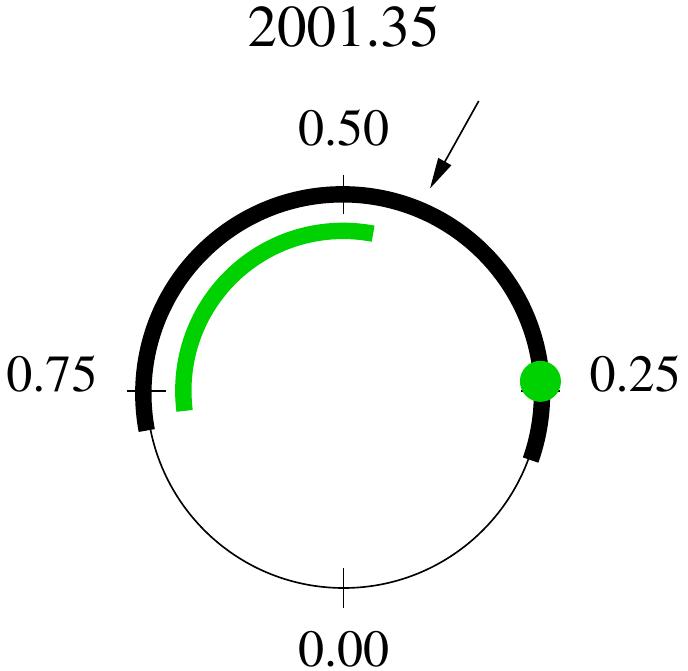}\\
\vspace{6mm}
 \includegraphics[width=0.16\textwidth]{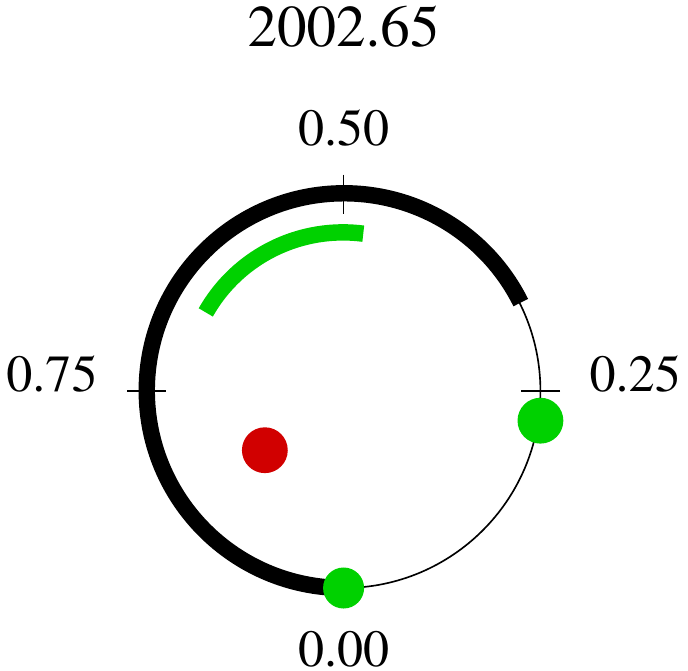}
 \includegraphics[width=0.16\textwidth]{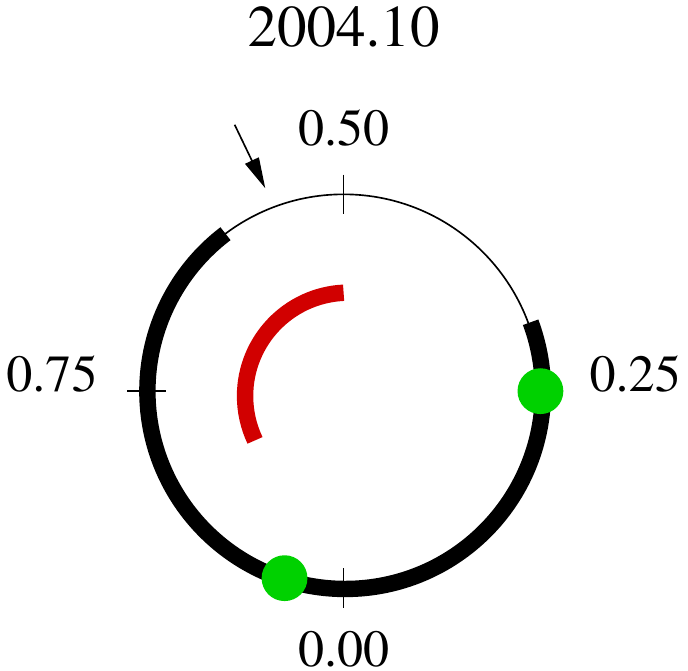}
 \includegraphics[width=0.16\textwidth]{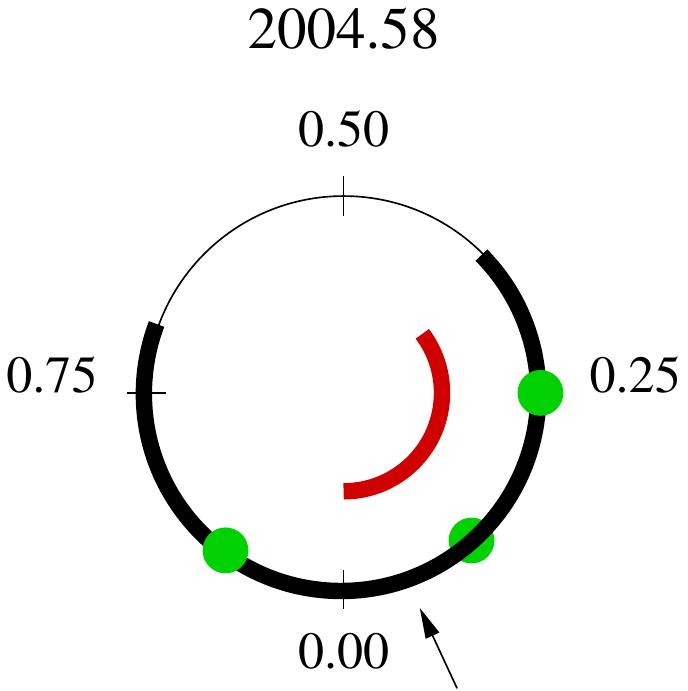}\\
\vspace{6mm}
 \includegraphics[width=0.16\textwidth]{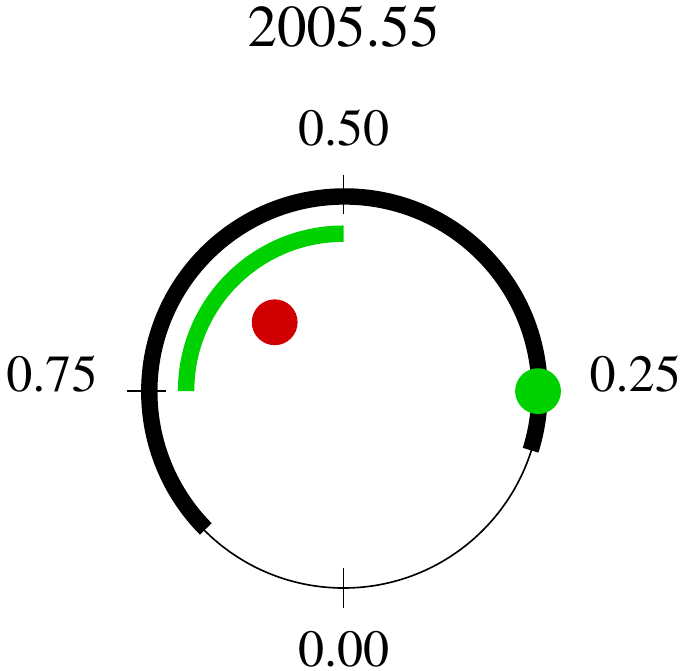}
 \includegraphics[width=0.16\textwidth]{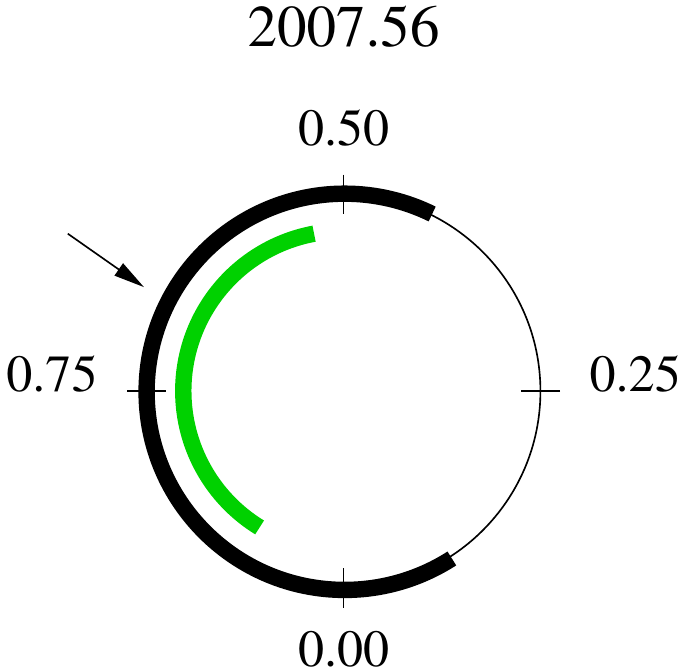} 
 \includegraphics[width=0.16\textwidth]{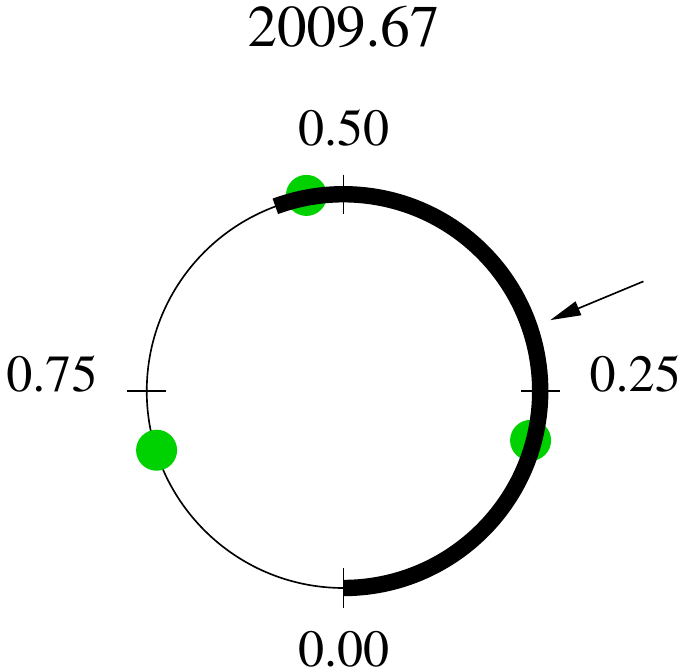} 

\caption{Possible connection between the dark photospheric spots and the chromospheric features. Approximate location of starspots from light curve inversion \citep{fkcom4} and from this paper (see Fig. \ref{fig:LI}) are marked with black arcs. Stronger spots from Doppler-images of \cite{fkcom5} and from this paper (see Fig. \ref{fig:maps}) are shown with green dots/arcs, green dots in the middle of the circles indicate polar caps. Phases of prominences (see Fig. \ref{fig:dynamic-spectra}), i.e., when the prominences cross the stellar disk are shown with black arrows outside the circles. Phases of increased \HA{} emission seen in Fig. \ref{fig:dynamic-spectra} are marked with red dots/arcs inside the circles.}
 \label{fig:drawing}
\end{figure}

\begin{figure}
 \centering
 \includegraphics[width=0.3\textwidth]{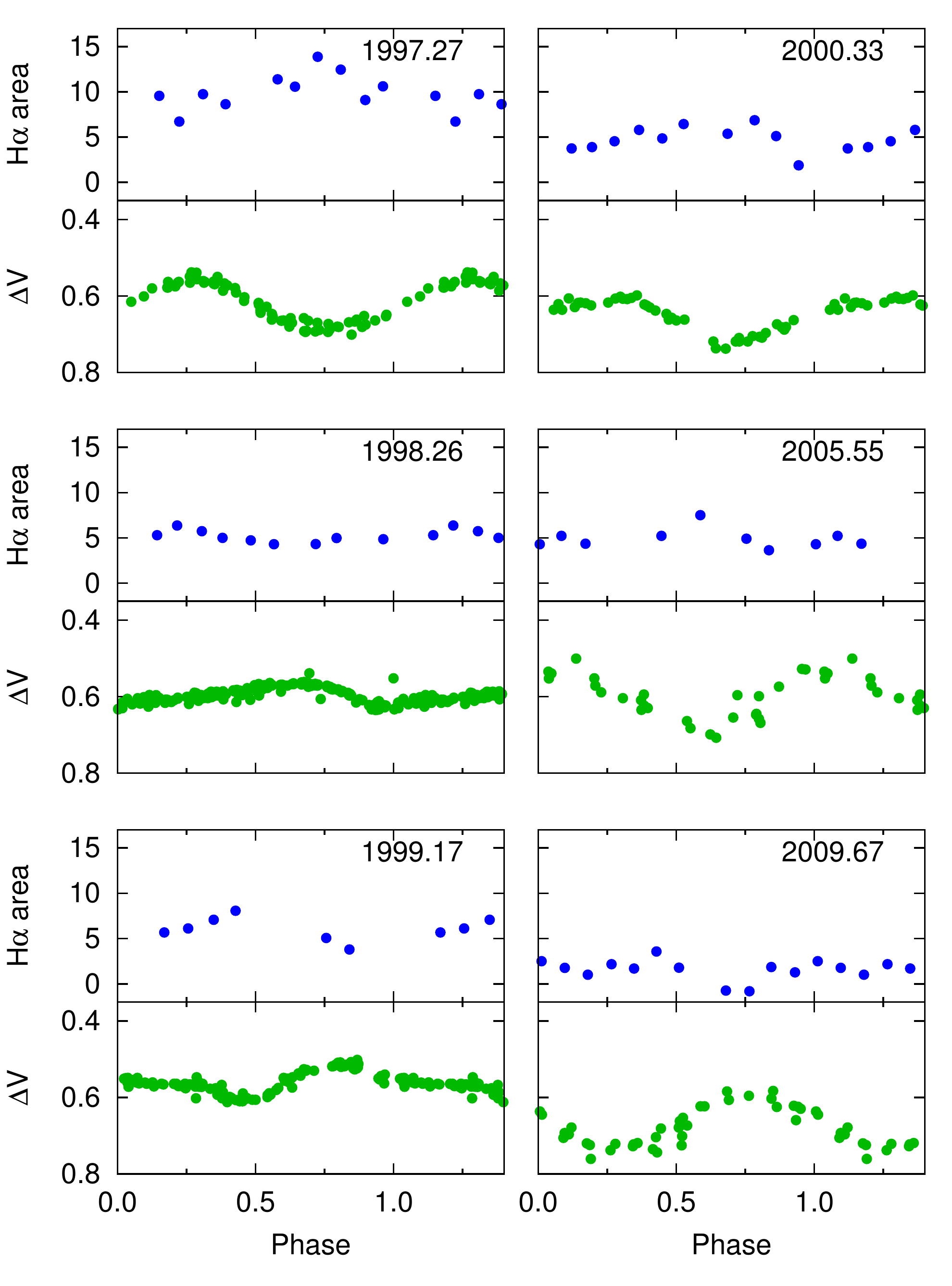}
 \caption{Comparison between the area of \HA{} residuals from Gaussian fits and light curves.}
 \label{fig:ha-lc-selection}
\end{figure}

The low-resolution \HA{} spectra from 1999 were observed between phases 0.35--0.40 on the first night and 0.75--0.78 on the second night. Although no changes could be observed during the nights, there is considerable difference between the spectra of the two nights. On the first night the \HA{} emission is stronger than on the second one. 
\cite{fkcom4} published light curves and light curve inversions of FK Com from 1979 to 2001. The light curve from 1999.21 is obtained just a few days after the low-resolution \HA{} spectra. In the light curve two spots can be seen: one at phases between 0.3--0.6, and one between 0.9--0.1. 
The Doppler map from 1999 April in \cite{fkcom5} also shows that the most spotted state is around phase 0.3, where the star showed high-latitude spots.
Thus the spectra of the first night show a spotted phase, while those from the second night show a less-spotted phase of FK Com. This can indicate a connection between the photospheric and chromospheric activity, as observed on the Sun and also on other stars: e.g.
on the BY Dra-type EY Dra (e.g., \citealt{2010AN....331..772K})
on the T Tauri-type TWA 6 \citep{2008MNRAS.385..708S};
on the K-dwarf LQ Hya \citep{2008A&A...481..229F}; 
and on RS CVn-type binaries, e.g. RT Lacertae \citep{2002A&A...388..298F};
and on the W UMa-type VW Cep \citep{1996A&A...313..532F}. 
The \HA{} emission is however  not necessarily connected to the photospheric spots: on SAO 51891 no modulation of \HA{} was observed, although changes caused by chromospheric inhomogeneities were clearly detected in the $V$ light curve, and in the Ca {\sc ii} HK, IRT and H$\varepsilon$ lines. In this case the \HA{} modulation might be possibly hidden by other activity signatures, e.g.  microflares \citep[see][]{2009A&A...499..579B}.

It is interesting to note, that the spectra taken at the spotted phase of the star have asymmetric shapes, with a stronger peak in the red side of the \HA{} line. 
\cite{2005A&A...434..221K} found that the violet peak of the \HA{} spectra shows a quasi-sinusoidal change with a maximum at phase 0.75 and a minimum at 0.25. The red peak was in maximum at phase 0.25 and in minimum at 0.75, but the changes were not so significant as at the violet one. During this time, the main spotted area could be seen between phases 0.5--1.0 \citep[see][]{2004AN....325..402K}, i.e. at the light curve minimum the violet peak was in maximum.

\subsection{Prominences}

In some cases in Fig. \ref{fig:dynamic-spectra} noticeable wave-like  features can be seen in the residual spectra. These are probably caused by prominences in the outer stellar atmosphere. According to the maximum amplitude of these waves the emission mostly originates from about one stellar radius above the surface.
In some cases of Fig. \ref{fig:dynamic-spectra} (1999.58, and possibly in 1999.41, 2001.43) multiple waves can be seen, possibly caused by prominence arcs of different size and location.
We compared these spectra to the photospheric features seen in the nearby light curve inversions \citep[][and this paper]{fkcom4} and Doppler-images \citep[][and this paper in Fig. \ref{fig:maps}]{fkcom5}. In Fig. \ref{fig:drawing} the phases of the spots, flares, and prominences can be seen.  We marked photospheric spot locations both from light curve inversions and from Doppler images (with black arcs and green arcs/dots). We determined the approximate phases of the prominences -- marked by black arrows outside the circle  --  by visual inspection of the dynamic \HA{} spectra shown in Fig. \ref{fig:dynamic-spectra}, selecting the phase when the Doppler shift of the prominence is zero. Sudden increases  (probably caused by flares) of \HA{} emission are also shown, marked by red arcs or dots.

In most cases, when the wave-like behaviour of the \HA{} region was observed, the prominence appeared at a spotted phase, except at 2004.10. In many cases, the phases of the prominences coincide with the increased level of \HA{} activity: in 1998.30, 2000.33,  2004.10, 2004.58.
At 1998.30, 1999.58, 2001.35, 2004.58, and 2007.56 the prominence is very close in phase to a dark spotted feature seen in the Doppler-map. These can be an indication that there is a connection between the prominence and the photometric spots as found by \cite{Huenemoerder}, who also observed in 1989 that the dark photometric spots and H$\alpha$ excess coincide in phase. 

A detailed description of the possible connection between the different activity features is given in Appendix \ref{sect:app:desc}.

\subsection{General level of activity}
\label{sect:general}
\begin{figure}
 \centering
 \includegraphics[width=0.35\textwidth]{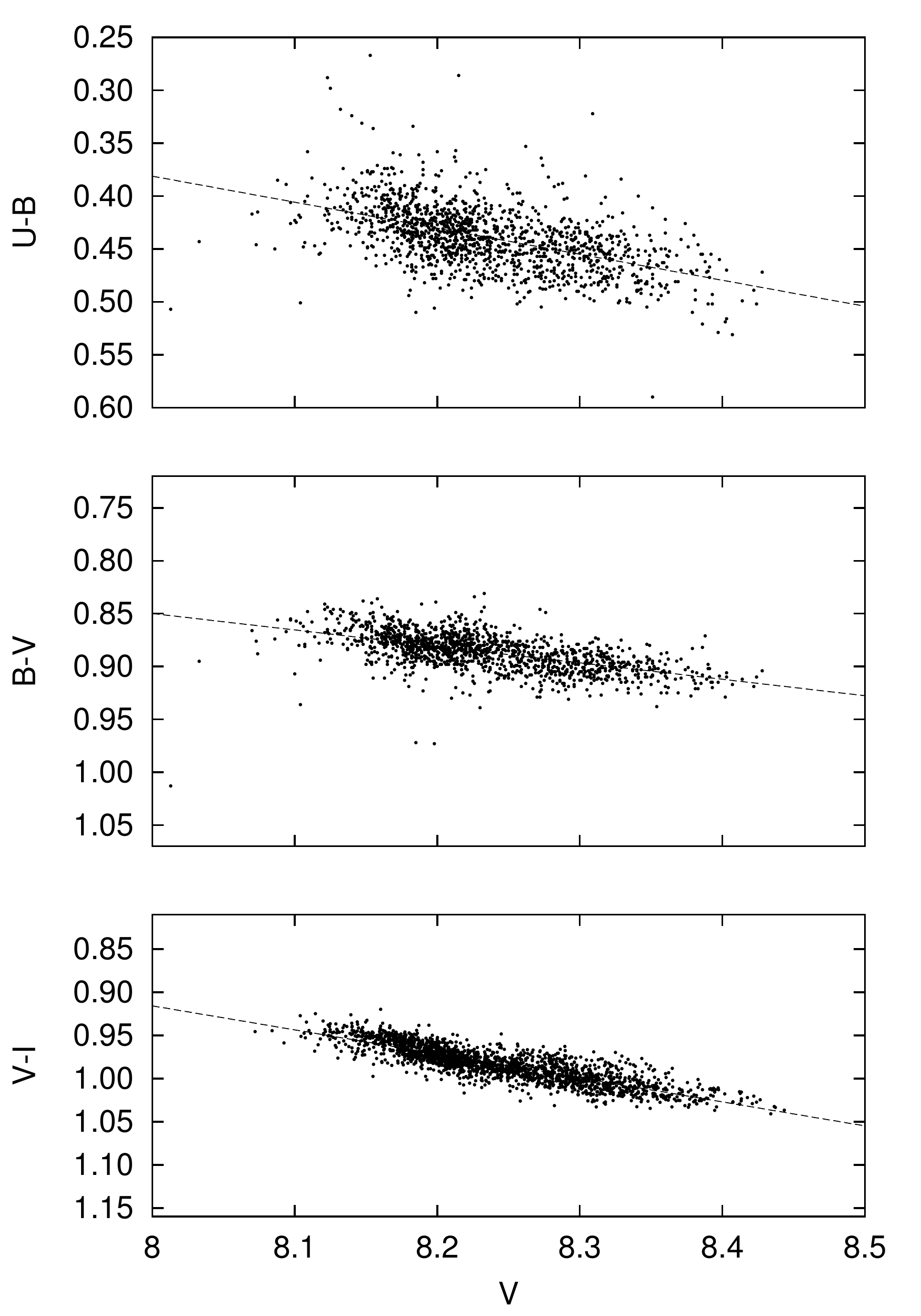}
 \caption{Magnitude--colour index curves of FK\,Com using photometric data of \cite{2013A&A...553A..40H} covering the years 1997--2010. Dashed lines show the best fit to the data.}
 \label{fig:CI}
\end{figure}

\begin{figure}
 \centering
 \includegraphics[width=0.40\textwidth]{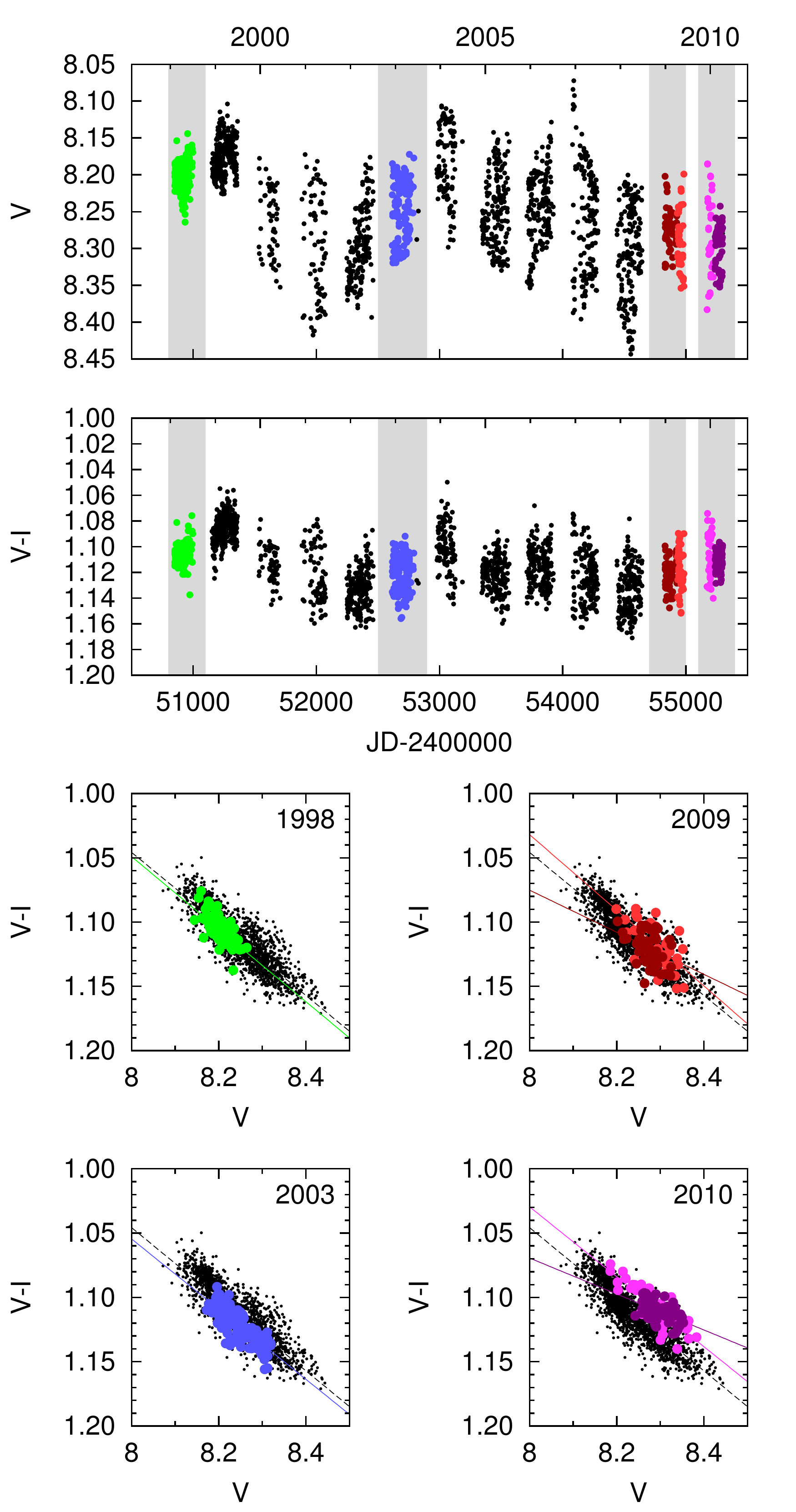}
 \caption{Light curve and magnitude--colour index curves of FK\,Com using photometric data of \cite{2013A&A...553A..40H} covering the years 1997--2010. The magnitude--colour index plots show four section of the light curve having lower amplitudes  plotted over all data, marked with small dots. Dashed lines show the best fit to the given sections.}
 \label{fig:CI2}
\end{figure}

%%%%%%%%% N E W   P A R T  %%%%%%%%
To measure the strength of the \HA{} line, we fitted  Gaussians to the residual of the \HA{} region after subtracting the spectrum showing the lowest activity, thus removing the effects of a possible disk present around the star.

Anti-correlated behaviour was observed between the $V$ light curve and \HA{} by \cite{Huenemoerder}, and this was the case in our low-resolution spectroscopic observations, too. If we compare phased light curves and the area of the residual Gaussians (see Fig. \ref{fig:ha-lc-selection}), we can find similar trends, although the relation here is much less clear, and was found only in some cases, while positive correlations and no \HA{} modulation have also been observed. One reason for this could be that the \HA{} emission is originating from both plage regions and prominences, and these features can be observed better at different phases. Namely, plage-like structures are better seen on the stellar disc, while prominences show themselves better when seen off-limb \citep[see e.g.][]{Hall:1992bx}. Note, that strict anticorrelation between the $V$ light curve and \HA{} is observed only when the spots and the \HA{} emitting regions are co-spatial.

Comparing the \HA{} intensities and the $\Delta V$ light curve (see Fig. \ref{fig:ha-strength}) we find, that they have a very similar trend between 2000 and 2006, although this connection cannot be seen in the earlier and later observations. The long-term positive correlation of the light curve and the \HA{} line (i.e., with a stronger chromospheric activity  as the star gets brighter) could be explained by facula-dominated activity, i.e., that most of the optical brightness changes are caused by hot spots -- facular regions, at least on the long-term base. The magnitude--colour diagrams of \cite{fkcom3} showed that activity of FK\,Com is dominated by spottedness and not photospheric facular regions: the star gets redder as it becomes fainter (see also \citealt{2008A&A...480..495M}). This finding can be confirmed by reconstructing these magnitude--colour diagrams using also the more recent photometric data of \cite{2013A&A...553A..40H} see Fig. \ref{fig:CI} and Fig. \ref{fig:CI2}. These diagrams though, show the cumulative effect of the rotational modulation and the long-term changes which can hardly be separated. 

\begin{figure}
 \centering
 \includegraphics[angle=-90,width=0.44\textwidth]{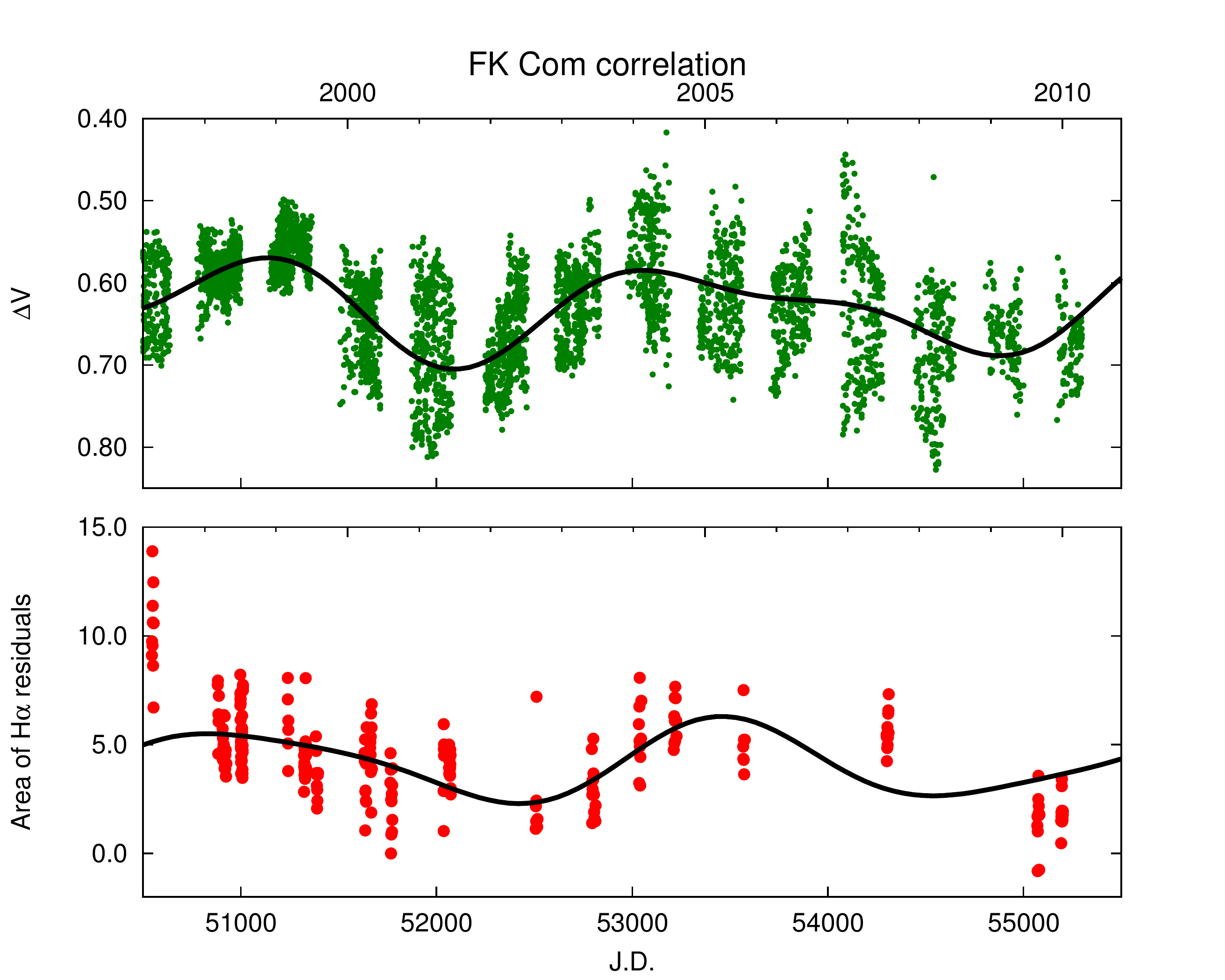}
 \caption{Top: $\Delta V$ photometry of FK\,Comae, bottom: \protect{\HA{}} line areas of
  the residual spectra. Continuous lines in the upper two plots show a two-period fit to the data (see text).}
 \label{fig:correl}
\end{figure}
\onlfig
{
\begin{figure}
 \centering
 \includegraphics[angle=0,width=0.44\textwidth]{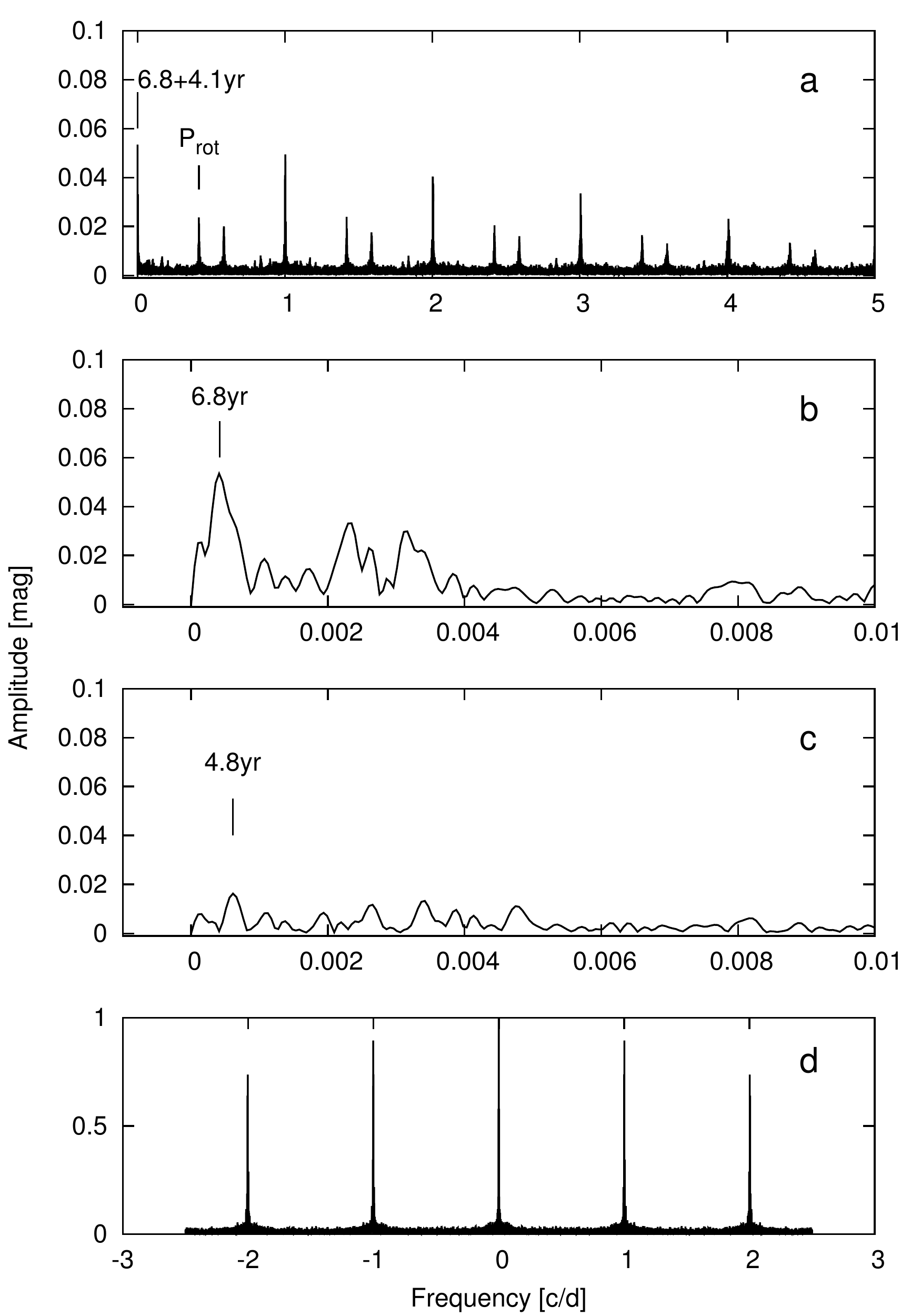}
 \caption{Fourier spectrum of the FK Com $V$ light curve in those year, where \protect{\HA{}} observations exist. {\it a)} shows the whole spectrum, {\it b)} is zoomed in to the long-period region. {\it c)} shows the zoomed-in spectrum after pre-whitening with the 6.8 year-long period. The bottom plot {\it d)} shows the spectral window.}
 \label{fig:fourier}
\end{figure}
}

In Fig. \ref{fig:correl}, a two-period fit to the $V$ data is drawn in the upper panel, with lengths of $6.8\pm0.5$ and $4.1\pm0.1$ years 
(the errors of the periods were estimated by the frequencies of the Fourier-spectrum 0 $.\!\!^{\rm m}$0005 below its maximum
which corresponds to 10\%
of the precision of the data used, cf. \citealt{error}.). 
A fit with the same periods is plotted on  
\HA{} area dataset, and describes it reasonably well, apart from the first data (1997) with some phase shift. These two periods (rather time-scales) were found from a simple Fourier analysis (see online Fig. \ref{fig:fourier}), and serve only to show the long-term changes in \HA{} relative to $V$. However, these time-scales are verified with a time--frequency analysis of the 32 years long $V$ light curve taken between 1979--2011, with the STFT (Short-time Fourier transform). The results are shown in the online Fig. \ref{fig:stft}. For the details of this method and its application see \cite{stft} and \cite{stft2}. After JD 2451000, i.e., around 1999 the long-term, mean brightness of the star shows an abrupt decrease. The cycle period is increasing from about 5.5 years in 1982 (2445000) to the present 8.3 years. In 1999 the cycle period was about 6.6 years and a constant other cycle showed up with an approximate length of 4.2 years, lasting till the present. This other cycle showed up after the flip-flop event in 1997--1998. Whether any change in the cycle periods appear after the recent (2009) event remains to be seen in 15--20 years.
\onlfig
{
    \begin{figure}[t]
    \centering
    \includegraphics[angle=0,width=0.5\textwidth]{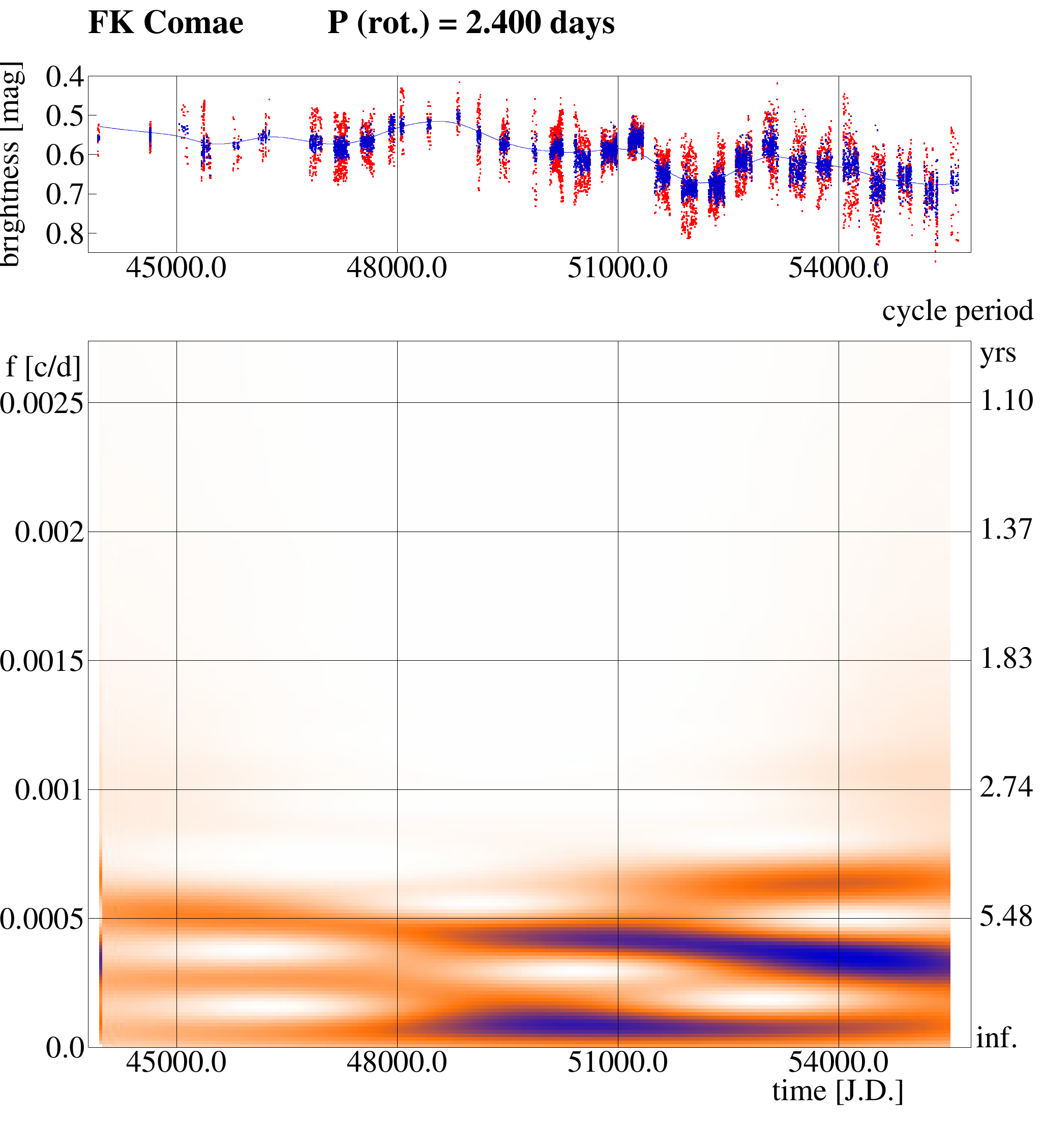}
    \caption{$\Delta V$ light curve (top) and its time--frequency analysis using the STFT (Short-time Fourier transform) method (bottom) showing multiple activity cycles of changing length on FK Comae.}
    \label{fig:stft}
    \end{figure}
}

The longer ($\approx$7 years) cycle dominates the \HA{} variations and seems to be shifted by about 1--2 years with respect to the brightness in $V$. With the growing spottedness between 1999--2002 the \HA{} excess decrease and spots dominate. From 2002 to 2004 the star brightens which means less spots, by 2004 the \HA{} is stronger, and after that more spots appear and the star becomes gradually fainter again. This could mean, that we see the same quasi-periodic behaviour in the photometry and the \HA{} spectra, i.e., the photosphere and the chromosphere, but the chromospheric changes are delayed with respect to the photosphere. 

\cite{1998ApJS..118..239R} investigated the long-term relations between the photospheric and chromospheric variations of 35 stars, including the Sun, using the Mount Wilson HK data (see e.g. \citealt{wilson}) with contemporaneous photometry from Lowell Observatory during the last decade until 1995. The results show that the younger stars of the sample, i.e., below 1--2~Gyr  age are fainter when their chromospheric emission increase while the older stars, including the Sun, show parallel behaviour showing brightening with higher chromospheric emission. Note, that these relations are not strict parallels or mirror images but show phase shifts and sometimes different patterns (see \citealt{1998ApJS..118..239R}, Fig. 3.). 
%Our star, FK~Com is a member of the HR~1614 moving group thus is about 2~Gyr old, or more, belonging to the old disk population \citep{2000A&A...357..153F}. Therefore, according to the results of \cite{1998ApJS..118..239R}, anticorrelation is suspected between the long-term photospheric and chromospheric changes, as observed.

%%%%%%%% END %%%%%%%%

It is also possible that cool spots are caused not just by magnetic activity but also vortices, as proposed by \cite{2011ApJ...742...34K}. In this case there would be no connection between the light curve and the \HA{} emission. Another possibility could be the structure of the spots: i.e. spot groups with mixed polarities vs. large unipolar spots, as these two kind of structures could cause very different \HA{} emission. In principle this could be verified by studying the magnetic field structure.

According to the Doppler images, bright regions are  present on the surface. Such hot regions were found on FK\,Com in Doppler images of \cite{fkcom5}, e.g. in 1993 June or  2003 June, or in some of the ones presented in this paper (see Fig. \ref{fig:maps}).  These bright spots tend to get more pronounced when the data quality is poor, and especially large phase gaps in the data result in hot spots in Doppler images, suggesting they are artefacts of the inversion method. This possibility is also supported by the fact that the bright regions  mostly occur near the cool spots -- this pairing of cool and bright features is a way for the inversion to mimic spectral features visible at certain phases but absent in others. This can also occur if there are significant changes in the spot configuration occurring during the observation run (see \citealt{2013A&A...553A..40H}). Therefore, their reality cannot be conclusively verified.

\begin{figure}
 \centering
 \includegraphics[width=0.40\textwidth]{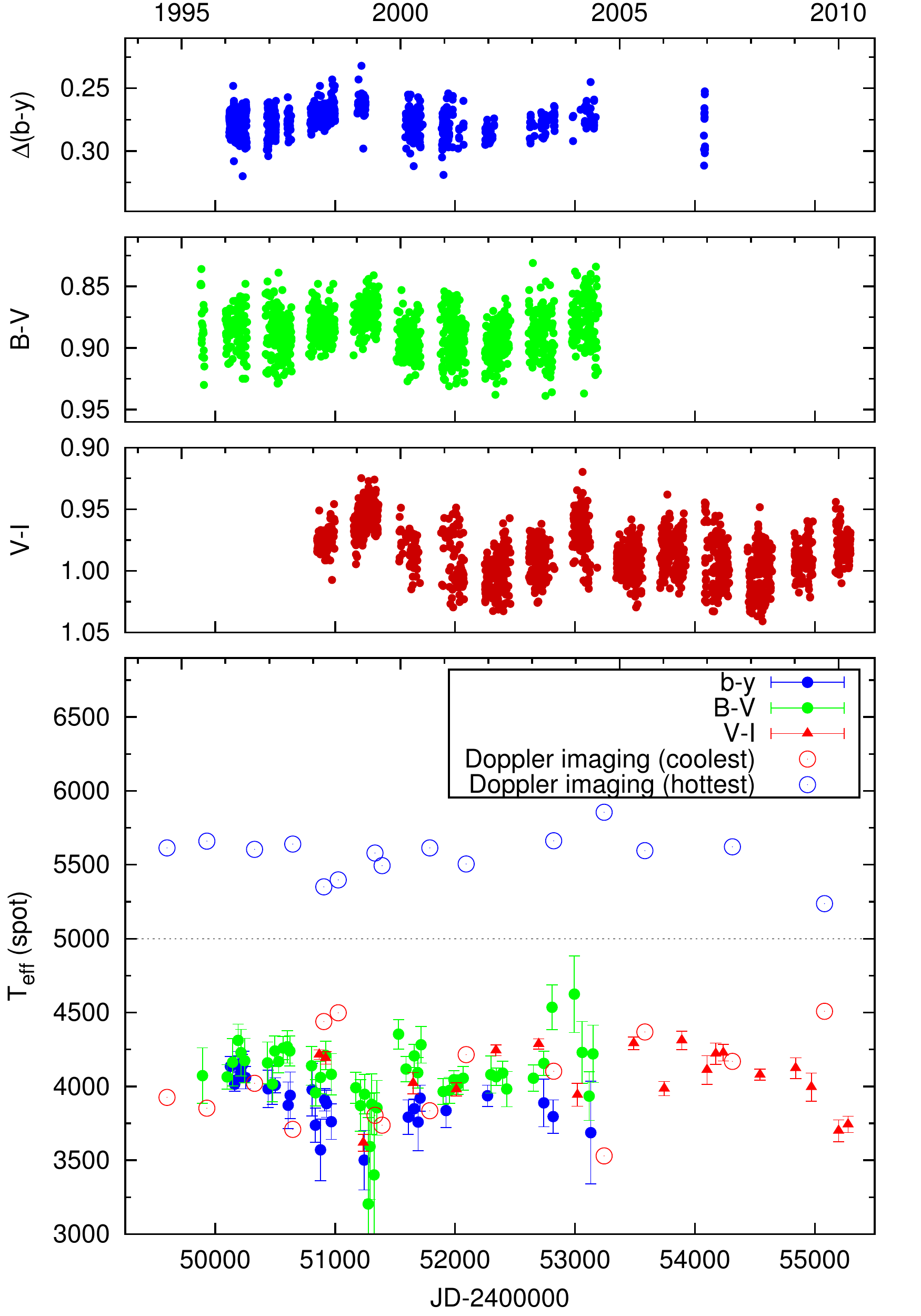}
 \caption{
The upper three panels show the long-term behaviour of the $b-y$, $B-V$ and $V-I_C$ colour indices which reflect the global temperature changes of the star. In the bottom panel the starspot temperatures using photometry and Doppler Imaging are plotted. The data points for the Doppler-map temperatures are from this paper, \cite{fkcom1}, \cite{fkcom2}, and \cite{fkcom5}. Horizontal line at 5000\,K shows the unspotted temperature.
}
 \label{fig:DIcontrast}
\end{figure}

In Fig. \ref{fig:DIcontrast} we plotted spot temperatures from both Doppler images and spot modelling.
Photometric spot temperatures were derived both from narrow- and wide-band data ($b-y$ and $B-V$,$V-I_C$ respectively), using {\sc SpotModel} \citep{sml}. The spot temperatures derived from the different colour indices generally match each other well. On average though, the spot temperatures from wide-band photometry are a little bit higher than from the narrow band data, which is probably due to more contribution of faculae to the wide-bands. The temperatures of the coolest areas from Doppler imaging agree fairly well with spot temperatures from photometry. The hottest regions have about 500\,K higher temperatures than the photosphere ($T_\mathrm{eff}$ = 5000\,K assumed). By 2010 the temperature of the coolest regions from Doppler imaging became higher supporting the measured increase (blueing) of the $V-I_C$ colour index.

The contrast of the Doppler images seems to decrease in time in the good-quality maps. In Fig. \ref{fig:DIcontrast}  minimum and maximum map temperatures in a uniform sample of Doppler images are given. All the maps used  for this plot are from SOFIN data and they have maximum phase gap of $<0.25$, i.e., cases where strong artefacts are not expected (temperatures are collected from this paper, \citealt{fkcom1}, \citealt{fkcom2}, and \citealt{fkcom5}). A linear fit to the coolest Doppler imaging temperatures in these maps shows that the spot temperature increases on average approximately 23~K per year during the time period 1994--2010. On the other hand, the hottest temperatures in the maps are typically around 5600~K (67\% of the time 5500--5700~K). The fit to the hottest temperatures exhibits a slight decrease over the years, but only at a level of approximately 4~K per year, and it cannot be considered significant. One has to note, though, that the unspotted surface of FK\,Com has a temperature of approximately 5000~K, and therefore the hottest temperatures are significantly above this.

The 2009/2010 observations are especially puzzling. The whole brightness level is low, the light curve has low amplitude modulation. As photometric observations show, a sudden phase jump or flip-flop occurred in the rotational modulation amounting about $130^\circ$  in longitude, sometime in 2009 March. The phase of the main minimum jumped from phase 0.6 to 0.3 as seen from Fig 6. The \HA{} spectral region, observed overlapping the photometric variations from late 2009/early 2010 shows very low level of chromospheric activity (see Fig. \ref{fig:ha-lc-selection}). In Fig. \ref{fig:dynamic-spectra} we can see that the level of \HA{} emission is indeed steady, it is originating from outside the stellar disk, and there is a wave-like structure indicating a prominence, that reaches the middle of the stellar disk at phase 0.2--0.4, coinciding with the light curve minimum. 

In the end of the 2009/2010 season by 2010 April, the amplitude of the rotational modulation became even lower and the main minimum shifted to phase 0.1. The spot temperatures from $V-I_C$ was steadily decreasing from 2005 to 2010. In 2009 August the lower spot temperature from Doppler imaging (DI), observed later than the phase jump (or flip-flop?) happened in 2009 March, is significantly warmer than that from the $V-I_C$ colour index. Similarly, in the second half of 1997 a flip-flop was observed by \cite{fkcom3}, and after that, in early 1998 the lower spot temperature from DI was also warmer than those from photometry (at that time we have temperatures from 3 colour indices). These two observations are suggestive for finding a correlation between flip-flop events and chromospheric activity on a long-term base.

To investigate further the relation of the bright and cool regions on FK\,Com, we compared magnitude--colour diagrams from different epochs. Four of these diagrams are shown in Fig. \ref{fig:CI2}. The $V$ vs. $V-I_C$ plots have very similar trends, but in the last two epochs this trend seems to be somewhat weaker (especially in the cases of lower amplitudes, plotted with a darker shade and with a separate fit). This can suggest, that there is a change going on in the activity of FK\,Com. This change might be connected with the results of \cite{2009MNRAS.395..282K} and \cite{Puzin:2014hj}, who studied longitudinal magnetic field on FK\,Com using low resolution spectropolarimetry. According to \cite{2009MNRAS.395..282K}, the average longitudinal magnetic field strength of FK\,Com was in the order of 200\,G in 2008 April, but \cite{Puzin:2014hj} found that in 2012 May there was no detectable magnetic field on the star. The measurements from 2011 April--May show significant detections of the longitudinal magnetic field, but with much reduced field strength compared to 2008 April (Korhonen et al. in preparation). The observations of \cite{Puzin:2014hj} are of lower accuracy than the ones used by \cite{2009MNRAS.395..282K}, and therefore, as the authors also hypothesise in their paper, the non-detection could be due to the field being too weak for them to detect. All this supports that the activity of FK\,Com has significantly decreased during the recent years: decreasing spot contrast of the Doppler maps, weak \HA{} levels in the last epochs, and decreasing longitudinal magnetic field strength.

Describing the complex phenomena observed on FK\,Com is not an easy task. According to the most likely explanation, the object is a coalesced contact binary system, possibly with a surrounding disk \citep{Webbink:1976db}, and often one or more prominence arcs. The stellar surface is covered both by bright plage regions, and dark starspots, as seen in the Doppler images. The spots in some cases can form a belt-like structure, resulting in an unvarying light curve that is fainter than the brightest state of the star (e.g. in 1993.31, see \citealt{fkcom4}) There is a connection between the changes in the chromosphere and the photosphere, but it is far from obvious. According to the later light curve amplitudes, polarimetric measurements spot contrast from Doppler images, the general activity level seems to decrease, but the average light curve level still indicates a spotted state. 

\section{Conclusions}

Based on the high and low resolution spectra presented in this work we draw the following conclusions:

\begin{itemize}
\item The \HA{} region shows a double-peaked emission, possibly indicating a circumstellar disc, and often prominence-like structures that reach to more than a stellar radius and are stable for weeks.  
\item The low-resolution \HA{} spectra show small-scale fast variability during one observing night on the time scale of hours, and the shape of the line profile changes between the nights more significantly. The first observing night, showing higher \HA{} emission level and a more asymmetric shape, coincided with the most spotted phase of the light curve and of the Doppler map;
\item Comparing the phased $\Delta V$ light curves and the excess \HA{} emission, we often find an anti-correlated behaviour of the two: the \HA{} emission peaks at the faintest state of the star: at the most spotted phase on the light curve and the Doppler maps;
\item This trend is not obvious on longer term: between 2000 and 2006 the $\Delta V$ light curve and the \HA{} emission seems to be correlated, but the observations before and after this time seem to behave differently;
\item The long-term correlation between the light curve and the \HA{} emission could be explained by the plage regions, but the magnitude--colour index in \cite{fkcom3} does not confirm this scenario: the star seems to get redder as it gets fainter, indicating a spot-dominated activity;
\item The connection between the \HA{} emission level and the light curve could be explained by the same quasi-periodic changes in the chromosphere and the photosphere, with the chromospheric changes being 1--2 years 'late' compared to the photosphere;
\item In the 2009/2010 observations the average level of the $\Delta V$ light curve is very low compared to low level of the \HA{} emission and the low contrast of the Doppler maps; 
\item On the whole, the activity of FK\,Com is unusually weak in 2009--2010. 
\end{itemize}

\begin{acknowledgements}
The authors would like to thank I. Tuominen, who unfortunately passed away in 2011 March, of his significant contribution to this work by starting collecting the unique dataset used in this work. The authors thank the anonymous referee for the helpful comments that improved the paper significantly.
K.V. and K.O. acknowledges support from the Hungarian Research Grant 
OTKA K-109276, OTKA K-113117 and the Lend\"ulet-2012 Program (LP2012-31)  of the Hungarian Academy of Sciences.
K.V. acknowledges support from the  Lend\"ulet-2009 grant, and the ESA PECS Contract No. 4000110889/14/NL/NDe.
K.V. was supported by the E\"otv\"os Scholarship of the Hungarian State.  K.V. acknowledges the hospitality of Finnish Centre for Astronomy with ESO (FINCA). Nordic Optical Telescope is operated on the island of La Palma jointly by Denmark, Finland, Iceland, Norway, and Sweden, in the Spanish Observatorio del Roque de los Muchachos of the Instituto de Astrofisica de Canarias. The data presented here were obtained in part with ALFOSC, which is provided by the Instituto de Astrofisica de Andalucia (IAA) under a joint agreement with the University of Copenhagen and NOTSA.

\end{acknowledgements}

\begin{appendix}
\section{Comparison between the different activity features}
\label{sect:app:desc}
In the following, we give a detailed description of the \HA{} activity in those cases where indications of a prominence was seen, and their possible connection with photospheric features:
\begin{description}
\item[1998.27/1998.30]
The prominence crossed the stellar disk at $\approx0.8-0.9$ phase. The 1998.27 set shows steady \HA{} emission, but there is a flare in the 1998.30 set at phase 0.9. The light curve shows a minimum around 0.75 phase. The Doppler image in \cite{fkcom5} show high-latitude spots at phases 0.34--0.52 and 0.78--1.00. The flare, the dark spots and the prominence coincide in phase.
\item[1999.41/1999.58]
In the 1999.41 observations we see a prominence at phase $\approx 0.2$. The light curve shows a minimum at phase 0.9 and a maximum at 0.4. The Doppler image from 1999 May shows a polar spot, and high-latitude spots at phases 0.6 and 1.0 and a southern spot at 0.97, but the one from 1999 July shows a much weaker spottedness around the same phases, and no polar spot. In this case the prominence is close to the brightest phase of the star. In both May and July there are brighter regions in the maps around 0.25 phase: in May we see two bright areas on the northern hemisphere, while in July it was found on the southern part.

1999.58 is the only case when we see a strong indication of two coinciding prominence loops in the \HA{} dynamic spectra, one around 0.5--0.6 phase and one at phase 1.0.
\item[2000.33]
We see a prominence crossing the stellar disk at 0.7 in phase. The light curve at this epoch shows a minimum at phase 0.7 and a maximum at 0.3. Doppler map from 2000 March shows high-latitude spots around phases 0.5--0.64, the one from 2000 April has dark features around 0.23, 0.52--0.66, 0.59--0.70, 0.78--0.84, 0.97 phases in different latitudes. This latter map was marked in the paper as less reliable. The spotted regions between phases 0.5 and 0.8 are present in both maps, one of these might be connected with the prominence.
\item[2001.35]
The light curve in this epoch shows a very symmetric sinusoidal shape with a minimum at 0.46 and a maximum at 0.97 in phase. In this case the wave-like \HA{} structure at phase $\approx 0.5$ in Fig. \ref{fig:dynamic-spectra} is less obvious, probably the prominence is hidden by the stellar disk around phase 1.0. This suggests a low-latitude feature. The Doppler map from 2001 May shows the strongest features -- both dark and bright regions -- around 0.25 phase, although this map was marked as less reliable. The following map from 2001 June shows two high-latitude spots at phases 0.20--0.38 and 0.54--0.74, an equatorial spot at 0.49, and a brighter feature around phase 0.5.
\item[2004.10]
The light curve in this epoch is a sinusoidal one with a minimum at 0.95 and a maximum at 0.43. The closest Doppler image is from 2004 July, shown in Fig. \ref{fig:maps} (in July the shape of the light curve is similar to that in February, but the amplitude is higher and the maximum is around phase 0.55). In this map the most pronounced features are two dark spots: a mid-latitude one around phase 0.1 and a high-latitude one between phases 0.2 and 0.3. 
The dynamic spectrum shows that the prominence is crossing the stellar disk at $\approx 0.5$ phase. The wave-like shape in the radial velocity is somewhat distorted, which could be a result of a mass ejection. We can see an increased level of \HA{} emission at phase 0.7, that could be originating from the prominence.
According to both the Doppler maps and the light curves in this case there is possibly no connection between the dark photospheric spots and the prominence. 
\item[2004.58]
The prominence crosses the stellar disk around phase 0.05, that is close in phase both to the light curve minimum and the large dark spots seen in the Doppler maps.
\item[2007.56]
The light curve shows a minimum at phase 0.8 and a maximum around phase 0.27. The Doppler map in Fig. \ref{fig:maps} is very interesting, it is dominated by both bright and dark spots between phases 0.55--1.00, with dark spots from the southern hemisphere reaching to higher latitudes, and a bright elongated feature at $\approx 45^\circ$ latitude. The dynamic spectrum shows a broad wave-like feature, indicating a prominence crossing the stellar disk around phase 0.6, close both to the light curve minimum and the spots found in the Doppler maps.
\item[2009.67] {The Doppler maps from 2009.67 and 2010.00 are quite similar: both have a spot group around phase 0.25 at relatively low latitudes. The temperature contrast is low, about 500\,K cooler than the unspotted surface in 2009 August and 600\,K as in 2010 January. There is no sign of a polar spot. There is no photometric observation during this 2009.67 period, but the light curve before these observations is sinusoidal one with a minimum at phase 0.32 and a maximum at 0.78. The following $\Delta V$ data shows similar behaviour as the previous one, with an additional secondary minimum around phase 0.65. 
The dynamic spectrum is very interesting: it shows an obvious wave-like structure, but the general level of the \HA{} emission is very low, barely higher than the lowest observed one. The source of the emission crosses the stellar disk at phase 0.3. 

}
\end{description}

\end{appendix}

\bibliographystyle{aa}

\bibliography{mn-jour,vida}
\end{document}